\documentclass[11pt,a4,english]{article}
\usepackage[T1]{fontenc}
\usepackage[latin9]{inputenc}
\usepackage{geometry}
\usepackage{verbatim}
\usepackage{amsmath}
\usepackage{graphicx}
\usepackage{longtable}
\usepackage{setspace}
\usepackage[round,sort,authoryear]{natbib}
\usepackage{amsfonts}
\usepackage{graphicx}
\usepackage{babel}
\usepackage{amsthm}
\usepackage[title]{appendix}
\usepackage{booktabs,caption}
\usepackage{enumitem}
\usepackage{amsfonts}
\usepackage{accents}
\usepackage{natbib}
\usepackage{bibentry}
\usepackage{amsmath}
\usepackage{mathtools}
\usepackage[colorlinks=,linkcolor=blue, citecolor=blue, ]{hyperref}
\usepackage{pdflscape}
\usepackage{pdflscape}
\usepackage{multirow}
\usepackage{comment}
\usepackage{mathtools}
\usepackage{lmodern}

\usepackage[T1]{fontenc}

\usepackage{upgreek}
\usepackage{titlesec}


\makeatletter
\newcommand{\nextverbatimspread}[1]{  \def\verbatim@font{    \linespread{#1}\normalfont\ttfamily    \gdef\verbatim@font{\normalfont\ttfamily}}}
\makeatother
\nobibliography*

\DeclarePairedDelimiter{\floor}{\lfloor}{\rfloor}
\usepackage[T1]{fontenc}

\theoremstyle{definition}
\newtheorem{theorem}{\normalfont\scshape \bfseries Theorem}
\newtheorem{definition}{\normalfont\scshape \bfseries Definition}
\newtheorem{assumption}{\normalfont\scshape \bfseries Assumption}
\newtheorem{lemma}{\normalfont\scshape \bfseries Lemma}
\newtheorem{example}{\normalfont\scshape \bfseries Example}

\newcommand\norm[1]{\left\lVert#1\right\rVert}

\numberwithin{equation}{section}

\AtBeginDocument{}

\newif\ifshow 
\showfalse 
\ifshow
  \includecomment{wrap}
\else
  \excludecomment{wrap} 
\fi

\titleformat*{\section}{\large \bfseries}
\titleformat*{\subsection}{\normalsize \bfseries}
\titleformat*{\subsubsection}{\normalsize \bfseries}

\begin{document}

\pagenumbering{roman}

\title{ {\Large \textbf{Sequential Break-Point Detection in Stationary Time Series: An Application to Monitoring Economic Indicators}\footnote{Article history: November 2016, November 2017, December 2021. A version of the current manuscript with Figures can be downloaded from $\textcolor{blue}{https://papers.ssrn.com/sol3/papers.cfm?abstract\_id=3983627.}$}
}
}
\author{\textbf{Christis Katsouris}\thanks{Ph.D. Candidate, Department of Economics, University of Southampton, Southampton, SO17 1BJ, UK. \textit{E-mail Address}: \textcolor{blue}{C.Katsouris@soton.ac.uk}.} 
\\

\\
First Draft: November 27, 2017
}

\date{November 27, 2021}

\maketitle

\begin{center}
\textbf{Abstract}
\end{center}
\vspace{-0.8\baselineskip}
Monitoring economic conditions and financial stability with an early warning system serves as a prevention mechanism for unexpected economic events. In this paper, we investigate the statistical performance of sequential break-point detectors for stationary time series regression models with extensive simulation experiments. We employ an online sequential scheme for monitoring economic indicators from the European as well as the American financial markets that span the period during the 2008 financial crisis. Our results show that the performance of these tests applied to stationary time series regressions such as the AR(1) as well as the AR(1)-GARCH(1,1) depend on the severity of the break as well as the location of the break-point within the out-of-sample period. Consequently, our study provides some useful insights to practitioners for sequential break-point detection in economic and financial conditions.
\\


\newpage 

\setcounter{page}{1}
\pagenumbering{arabic}

\newpage

\section{Introduction}

Monitoring financial stability using break-point detection schemes provides a mechanism for avoiding economic losses. From the risk management point of view, structural instability in time series regression models translates into presence of systemic risk in financial markets. According to \cite{gramlich2011structural} these vulnerabilities make the financial systems less resilient and risk-absorbent. On the other hand, current innovations in financial products increase the monitoring complexity in capturing dynamic systemic risk transmission. From the economics point of view,  detecting structural breaks in economic conditions and financial stress indicators provides a signal for an upcoming recession allowing policy-makers to assess macroeconomic conditions in real time and thus decide whether a warning should be issued. 

Specifically, such tail events are captured by early warning indicators, (EWIs), which are used by financial institutions and policymakers as a macro-prudential tool and a signalling tool for economic recessions. Therefore, constructing EWIs permits to sequentially monitor economic conditions (ECIs) and financial stability indicators (FSIs) as well as the long-term equilibrium of economic fundamentals signalling this way an upcoming recession when these deviate from predefined thresholds. In this field, two econometricians Mark W. Watson and James Stock have pioneered the development and monitoring of such indicators. The study of \cite{stock1989new} discusses relevant aspects for developing leading economic indicators (such as variable selection methods) for monitoring the US economy during the late 1980s. Furthermore, the study of \cite{watson1991using} discusses the use the probit and logit models for predicting the probability of recession (see, also \cite{estrella1998predicting}). 

Furthermore, according to \citet{babecky2012banking} monitoring suitable EWIs provides a tool for optimum decision making for necessary macroprudential measures aimed at reducing the risk of occurrence of financial crises by mitigating their impact on the economy. More importantly, the classification of appropriate EWIs depends on: (i) the definition of crisis occurrence\footnote{More specifically, an economic crisis is defined as the beginning in the quarter in which the real GDP of an economy falls below the preceding 4-quarter moving average and ending in the quarter in which real GDP reaches the pre-crisis level (see \cite{vegh2014road}).}, (ii) the nature of the crisis (banking, currency, financial, economic) and (iii) the characteristics of the crisis on the real economy.  A currency crisis can include high exchange rate depreciation, a banking crisis might include the exposure of banks to systemic risk while a debt crisis might include the need for a debt restructuring.  Therefore an aggregated collection of time series observations from such financial sectors contributes to the robust construction of economic indices for monitoring occurrence of economic crises \citep{babecky2012banking}. 

Our study considers both linear time series regressions such as the stationary AR(1) model as well as the non-linear AR(1)-GARCH(1,1) model for sequentially detecting structural breaks using weekly and monthly frequency NFCI and CISS indices. We consider first differenced stationary series following the literature of structural break testing procedures which operate under the assumption of stationarity and ergodicity. Our aim is to investigate the properties of various break-point detectors when these are employed within the sequentially monitoring framework of \cite{chu1996monitoring}. Furthermore, for statistical inference purposes we consider the hitting times of Brownian motions (stopping rule) since the underline stochastic processes of the proposed test statistics are assumed to be mean-reverting. Moreover, we investigate the finite and large sample performance of the break-point detectors via extensive Monte Carlo simulation experiments in which we obtain the empirical size, power function and average run length distributions for a fixed nominal size (see, also \cite{aue2009delay}).

\newpage 

\subsection{Literature Review}

In this study we distinguish between the types of break-point detection procedures, that is, the in-sample (historical-based) procedures and the sequential monitoring procedures. The historical monitoring procedures include the various F-tests which are suitable for testing the null hypothesis of no structural instability based on the full sample such as the studies of \cite{chow1960tests}, \cite{dufour1982generalized} and \cite{hansen2001new} who test for structural breaks in the US Labour productivity using the Chow statistic. Similar frameworks which operate under the assumption of a known break-point location include the residual-based tests for detecting parameter constancy of regression coefficients under the null hypothesis. In particular, \cite{brown1975techniques} propose the OLS-CUSUM test based on cumulated sums of recursive residuals for testing for single structural change in the model coefficients of the linear regression model. 

Further studies related to historical monitoring procedures include test statistics based on recursive residuals (see, \cite{kramer1991recursive}) as well as refinements of CUSUM type test statistics based on OLS-residuals such as \cite{kramer1988testing}, \cite{ploberger1990local} and \cite{ploberger1992cusum}, \cite{deng2008limit} and \cite{andreou2008restoring} who propose local power corrections. Moreover, the seminal study of \cite{Andrews1993tests} propose likelihood ratio and Wald-type statistics for testing for testing the null hypothesis of no parameter instability in linear regression models at unknown break-point locations. Lastly, \cite{leisch2000monitoring} extend the aforementioned methodologies to the generalized fluctuation test. 

In-sample structural break-testing procedures were extended to  testing procedures using an out-of-sample monitoring scheme; implying an estimation window of fixed size and an out-of-sample window for testing. In particular, in this direction the studies of \cite{kuan1994implementing} and \cite{chu1995moving} propose the recursive estimates  and moving estimates tests for parameter stability respectively. Additionally, \cite{chu1995mosum} propose the OLS-MOSUM test which is constructed based on sums of a fixed number of residuals from a rolling window. More recent applications which include comparisons between historical and online monitoring include \cite{zeileis2010testing} where the authors use M-type statistics to monitor structural breaks in the Chinese and Indian exchange rate regimes and \cite{zeileis2005monitoring} with focus on sequential monitoring procedures in dynamic models. 

Sequential monitoring frameworks are investigated in the studies of \cite{berkes2004sequential} and \cite{aue2013structural}. In practise, within a sequential monitoring scheme model-based estimates and functionals from 
continuously updated time series observations are compared to the corresponding estimates based on observations from the historical period. Under the law of iterated logarithms a large repetition would imply to incorrectly reject the null hypothesis of no structural break with probability approaching to one which ensure that the monotonically increasing property of the power function of these tests holds. On the other hand, in-sample structural break testing procedures are based on controlled asymptotic size properties which can satisfy a uniform property under further regularity conditions (see, \cite{chu1996monitoring} and \cite{leisch2000monitoring}). Generally, current methodologies in the literature, regardless of the monitoring scheme employed, focus on structural change testing and estimation for distribution moments which implies break detection in the conditional mean, the conditional variance or higher moments of partial sum processes (\cite{hansen2000testing}, \cite{horvath2001empirical}, \cite{kulperger2005high} and \cite{pitarakis2004least}). Other studies focus on developing methodologies for testing for multiple break-points such as in the studies of \cite{bai1997estimating}, \cite{bai1998estimating} and \cite{andreou2006monitoring}, although multiple break-point testing is not so common for sequential monitoring schemes. We leave this aspect for future research.

\newpage 

\subsection{Motivation and Outline}

In this paper, we focus on sequential break-point detection for linear and non-linear time series regression models. In particular, we focus on detecting for structural breaks based on linear time series models such as the stationary AR(1) regression model using time series observations of economic condition indices. More precisely, we utilize regression estimate-based processes as well as regression residual-based processes for the construction of the testing framework. The former type of fluctuation processes includes tests such as the recursive-estimates process, (RE)  and the moving estimates test, (ME). The latter includes recursive residuals based tests such as the CUSUM test and the MOSUM test or OLS residuals-based tests such as the OLS-CUSUM test and the OLS-MOSUM test \citep{chen2002tests}. We evaluate the asymptotic properties of these test statistics with a Monte Carlo simulation study where the presence of empirical size distortions or the non-monotonicity of power can indicate for possible problematic assumptions. \cite{kuan1994implementing} consider the ME and FL processes and suggest modified versions to deal with the problem of size distortion for variables with high autocorrelation.   
  
The paper is organized as follows. Section \ref{Section2} presents the econometric model and the sequential break-point detection framework along with main modelling assumptions. Section \ref{Section3} presents the main findings of the Monte Carlo simulation study based on the break-point detectors we examine in this paper. Section \ref{Section4} presents a discussion on the sequential monitoring procedure. Moreover, a discussion related to momentary and economic policy is also presented. Section \ref{Section5} summarizes the main concluding remarks and discusses relevant aspects for future research.

\section{Econometric Model and Sequential Break-Point Detection}
\label{Section2}

\subsection{Modelling Assumptions}

Consider the standard linear regression model 
\begin{align}
\label{eq:model}
y_t = x_{t-1}^{\prime} \beta_t + \epsilon_t, \ \ \ \ \ \ \  t \in \{1,...,n,n+1,..., N\}
\end{align}
The regression model given by \eqref{eq:model} corresponds to observations at times $\{t_i\}_{i=1}^{N}$, with $y_t$ the response variable at time t, $x_{t-1} = \big(1 , x_{2,t-1},..., x_{p,t-1} \big)^{\prime}$ the vector of $p$ explanatory variables and $\beta_t$ the model coefficients.  Define the OLS residuals and the OLS estimator $\hat{\beta}^{(n)}$ respectively based on the first $n$ observations such that
\begin{align}
\hat{\epsilon}^{(n)}_t = y_t - x_{t-1}^{\prime} \hat{\beta}_t^{(n)}, \ \text{and} \ \  \hat{\beta}_t^{(n)}= \left( \sum_{t=1}^n x_{t-1} x_{t-1}^{\prime} \right)^{-1} \left( \sum_{t=1}^n x_{t-1} y_t \right)  
\end{align}
A consistent estimator for the variance of the residuals is $\hat{\sigma}^2 = \frac{1}{n-p} \sum_{i=1}^{n} \hat{\epsilon}_t^{(n)}$. 

Furthermore, the historical period consists of the time series observations $1 \leq t \leq n$ while the monitoring period includes the time series observations  $n+1 \leq t \leq N$, where $N$ can be a multiple of the historical period such that $N = nT$ with $T = \left\{ 1,2,.... \right\}$. Then, we can formulate the null hypothesis of no parameter instability within the monitoring period and the alternative hypothesis of a single structural break at an unknown location within the monitoring period based on the following assumptions presented in the framework of \cite{chu1996monitoring}.

\newpage

\begin{assumption}
\label{Assumption1}
(Non-contamination) It corresponds to a historical period with sample size $n$ such that the parameter vector includes stable coefficients such that 
\begin{align}
H_0 : \beta_t = \beta_0 \ , \ t=\{1,...,n\}
\label{non-contam}
\end{align}
\end{assumption}
\begin{assumption}
\label{Assumption2}
(Contamination) During the monitoring period the null hypothesis corresponds to no structural break versus structural instability at some post-historical time $c^{*}$ 
\begin{align}
H_0 &: \beta_{ t } = \beta_0 \ , \ t= \{n+1,...,N\} 
\\ 
H_1 &: \beta_{ t_1 } = \beta_1,\ t_1=\{n+1,...,n + \kappa\} \ \text{and}  \ \beta_{ t_2 } = \beta_2 \ , t_2=\{n + \kappa+1,...,N\} , \beta_1 \neq \beta_2. 
\end{align}
\end{assumption}
The parameters $\beta_1$, $\beta_2$ and $\kappa$,  where $\kappa > 1$ is the break-point location are the theoretical model parameters which are to be estimated from the fitted econometrics model under examination. In particular, for the Monte Carlo Simulation Study of the paper we consider a stationary AR(1) model with an intercept such that $y_t = \mu + \rho y_{t-1} + \epsilon_t$. However, in this section we consider a general linear regression model such that $y_t = x_{t-1}^{\prime} \beta_0 + \epsilon_t$ for $t = \left\{ n + 1,..., N \right\}$.  
\begin{definition}
(Monitoring Scheme). A monitoring scheme is a stopping time, which indicates the detection of a structural instability in the econometric model. Let $\Gamma \big( n,t \big)$ be the detector function and $b(t)$ the boundary function over an appropriate common time interval (monitoring period). The empirical fluctuation process rejects the null hypothesis, $H_0$, when the detector becomes greater than the boundary function at a given time within the monitoring period.
\end{definition}

\begin{assumption}(Finiteness of Moments and FCLT)
\label{Assumption3}
\begin{enumerate}
\item[A1.] $\{\epsilon_t\}_{t=1}^{N}$ is a homoskedastic martingale difference sequence with $E[\epsilon_t^2  ]=\sigma^2$,
\item[A2.] $\left\{ x_{t-1} \right\}_{t=1}^{N}$ has at least a finite second moment, i.e. $\lim$ $sup_{n \to \infty} \sum_{t=1}^{n} \norm{ x_{t-1} }^{2 + \delta} < \infty$
where $\delta>0$ and $\displaystyle{ \lim_{N \to \infty} \frac{1}{N} \sum \limits_{t=1}^{N} x_{t-1} x_{t-1}^{\prime} \overset{p}{\to} \mathbf{\Omega} }$ for some finite non stochastic matrix $\Omega$.
\item[A3.] Under A1 and A2, $\left\{ x_{t-1} \right\}_{t=1}^{N}$ satisfies the Functional Central Limit theorem (FCLT) 
\begin{align}
S_N(r) 
= 
\left( N^{-1/2} \mathbf{\Omega}^{-1/2} \sum_{t=1}^{ \floor{ N r } } x_{t-1} \epsilon_t \right) \overset{d}{\to} \mathbf{W}( r ), \ 0 < r < 1. \label{fclt}
\end{align}
\end{enumerate}
\end{assumption}
and assume $\bigl| \sum_{t=1}^n \epsilon_t \bigr| = o_p(\sqrt{n})$,  as $n \to \infty$. 

Based on the formulation of the testing hypothesis and the definition of the monitoring scheme, detecting for structural break in the model implies rejecting the null hypothesis of no structural instability when the detector has high enough fluctuations to cross a pre-specified boundary approximated with the crossing probabilities of the Brownian motion. Similar assumptions can be found in the  generalized fluctuation testing framework of \cite{leisch2000monitoring} and  \cite{zeileis2005monitoring} as well as in the study of \cite{andreou2006monitoring}). Furthermore, the error term $\{\epsilon_t \}_{t=1}^{N}$ is assumed to be exogenous (representing economic, monetary and financial market shocks) but an \textit{i.i.d} Gaussian process. Furthermore, we consider that the FCLT result holds which implies a weak convergence for the partial sum processes of the model residuals and thus ensures that the asympototic behaviour of a sequential test $\{\theta_n \}$ is similar to the probability one convergence; i.e., spend nearly all of its time arbitrarily close to the same limit process (see, \cite{kushner2003stochastic}).

\newpage 

\subsection{Sequential Break-Point Detection Framework}
\label{Sec:Seq Mon Fra}

The sequential (online) break-point detection procedure\footnote{Notice that the sequential statistical analysis was first introduced during the 1940s with the purpose of increasing efficiency in acceptance sampling and extended to survival analysis models such as the proportional hazard model \citep{siegmund1986boundary}.} is briefly explained with the following steps. First, a random process produces randomly the observations $x_t$ which are received in the algorithm sequentially (online). Second, the detector function based on a pre-specified rule (boundary) decides on a timely manner (stopping rule) whether a break has occurred at an unknown break-point location within the monitoring period. Moreover, relevant probability bounds ensures that the trade-off between high volume of false-alarms and low volume of true positives is accommodated in the sequential procedure (see,  \citep{brodsky2013nonparametric}). Furthermore, the limiting theorems of sequential theory and Brownian motion provide the necessary tools for constructing a sequential monitoring framework.   
\begin{definition}(Stopping time)
A stopping time of a sequential monitoring process is  
\begin{align}
\tau_n(\kappa) = \mathsf{min} \ \bigg\{ \ \kappa \geq 1 : \Gamma \big( n=n_0, t \big) \big|_{t=\kappa} \ \geq \ b \left( n=n_0,t \right) \big|_{t=\kappa} \ \bigg\}
\end{align}
where the detector function and boundary function are chosen so that under Assumption \ref{Assumption2}, the following probability limits hold 
\begin{align}
\underset{ n \to \infty }{ \mathsf{lim} } \mathbb{P}_{H_0} \bigg[ \tau_n(\kappa)  < \infty \bigg] = \alpha  \ \ \ \text{and} \ \ \ \underset{ n \to \infty }{ \mathsf{lim} } \mathbb{P}_{H_1} \bigg[ \tau_n(\kappa)  < \infty \bigg] = 1 
\end{align}
where $\alpha$ is the significance level (nominal size of the null hypothesis) such that $0< \alpha < 1$.
\end{definition}

More specifically, the detector function is a function of the historical period and is updated in a sequential manner. Furthermore, the stopping time (or decision rule) as well as the boundary function requires a critical value in order to detect parameter instability (see Section \ref{boundaries}), which can be also seen in terms of a likelihood ratio procedure, such that 
\begin{align}
\tau_n(\kappa) = \mathsf{inf} \left\{ \kappa \geq 1 : \frac{ \Gamma \left( n=n_0, t \right) \big|_{t=\kappa}}{ b(n=n_0,t)\big|_{t=\kappa} } \geq \lambda_n(\alpha) \right\}  
\end{align}
Therefore, based on the above sequential monitoring scheme and suppose that Assumptions \ref{Assumption1}-\ref{Assumption3} hold, then we can test for a structural break in the coefficients of the econometric model within the monitoring period. In terms of the relevant methodologies currently presented in the literature,  \cite{leisch2000monitoring} propose an econometric framework for sequential monitoring for structural breaks using recursive and moving estimates processes, which are cumulative sums of model residuals. \cite{horvath2004monitoring} consider the sequential monitoring of the mean function using the OLS and the recursive CUSUM of residuals\footnote{Notice that the standard CUSUM process is based on recursive model residuals and has been proposed in the seminal study of \cite{kramer1991recursive}.}, where emphasis is given in the properties of the boundary function (i.e.,$b_6(t))$ in regards to the break-point location. Furthermore, based on a similar monitoring scheme, \cite{huvskova2010simple} propose a detection in the distribution of sequential arriving time series observations while \cite{chochola2008sequential} consider an OLS-CUSUM square process to monitor sequentially changes in the variance, a method also presented by \cite{chen2010sequentialmon}. Lastly, \cite{zeileis2005monitoring} consider both classes of processes (estimates-based and residual-based) using  econometric models and emphasis on rescaling the parameters when serial correlation exists, which is closer to the framework of our study.

\newpage 

\subsubsection{Sequential Estimates-based processes}

Let $\{t_i\}_{i=n+1}^{i=N}$ be the stationary time series observations within the monitoring period. Denote with $[Nq]$ to represent the discrete-time location (i.e., $[Nq] \mapsto N + k, q \in [0,1])$ and $h \in [0,1]$. Then, the $j-$th recursive and moving estimates of the parameter $\beta$ for the econometric the model \eqref{eq:model} are given by the expressions below (see e.g., \cite{kuan1994implementing})
\begin{align}
\hat{\beta}_{j}
&= 
\left( \sum_{t=1}^j x_{t-1} x_{t-1}^{\prime} \right)^{-1} \left( \sum_{t=1}^j x_{t-1} y_t \right), \ \ \ \text{for} \ \ j={n+1,...,N}
\\
\hat{\beta}_{j}^{(h)}
&= 
\left( \sum_{t=j+1}^{j+[Nh]} x_{t-1} x_{t-1}^{\prime} \right)^{-1} \left( \sum_{t=j+1}^{j+[Nh]} x_{t-1} y_t \right), \ \ \ \text{for} \ \ j={0,...,N-[Nh]+1},  
\end{align}
where
\begin{align}
\mathbf{\Omega}_{[Nq]}= \frac{1}{[Nq]} \sum_{t=1}^{[Nq]} x_{t-1} x_{t-1}^{\prime} \ \ \ \text{and} \ \ \ \mathbf{\Omega}_{[Nq]}^{(h)} = \frac{1}{[Nh]} \sum_{t=[Nq]+1}^{[Nq]+[Nh]} x_{t-1} x_{t-1}^{\prime}
\end{align}  
    
\paragraph{Recursive Estimates process:}
The process as described by \citet{chu1996monitoring} and by \cite{leisch2000monitoring}  is based on recursive estimates,(RE), for detecting any structural instability in the model. The RE detector is defined as the deviation of the updated parameter estimate $\hat{\beta}_{(m)}$ from the historical parameter estimate $\hat{\beta}_{(n)}$. Under the null hypothesis the process stays in control since there is no structural break and the detector is below the boundary function. Then the RE detector is given by
\begin{align}
R_n( t ) = \frac{m}{\hat{\sigma} \sqrt{n}} . {\mathbf{\Omega}}_{(n)}^{1/2} \left[ \hat{\beta}_t^{(m)} - \hat{\beta}_t^{(n)}    \right],  \ \mathbf{\Omega}_{(n)}= \frac{1}{n}\mathbf{X}_{(n)}^{T} \mathbf{X}_{(n)} , \ m = \floor{ k + t(n-k)}
\end{align}
where  the lower integer (floor) part of a number. Let  be the norm of a vector or matrix, that is, $ = \sum_{i=1}^{n} u^2_i  $. We then can define the stopping time $\tau_m$
\begin{align}
\tau_m = \mathsf{min} \left\{ \kappa : \norm{ \sum_{m< t < m+\kappa}  F_n(t)  } > b(t) \right\}
\end{align}
The RE process with a commonly used boundary $b_1(t)$ (see equation \ref{boundary 1}) and with time rescaling i.e., $t \mapsto k/n$ (invariance principle of Brownian motion) can be approximated by the Brownian bridge limit with a finite probability bound such that 
\begin{align}
\lim_{n \to \infty} \mathbb{P} \bigg[ S_n \geq \sqrt{n}A(k/n, \delta) \bigg] 
=
\mathbb{P} \bigg[ W(t) \geq A \big( t,\delta \big) \bigg]
\end{align}
More detailed regarding the properties of these probability limits can be found in the study of  \cite{robbins1970boundary} as well as in Corollary 3.6 of \cite{chu1996monitoring}). Moreover, the particular probability bounds permit to monitor the fluctuations induced by the empirical fluctuation process sequentially, which ensures the existence of a crossing probability with a controlled asymptotic size. Although, due to the monitoring scheme which implies continuously updated comparisons of historical based estimates and out-of-sample based estimates results to a higher computational complexity (and rate of convergence) than conventional processes.

\newpage 

\paragraph{Moving Estimates process:} This process presented in \cite{leisch2000monitoring} consists of moving sums of estimated regression coefficients, which compares sequentially the deviation of the moving-window estimates to the whole sample estimates for break point detection. The particular process seems to be more sensitive to the choice of the boundary function, to the sample size of the historical period as well as to any violations in the independent assumption of observations. To deal with the problem of serial and high autocorrelation in stationary time series regression models (e.g., models we use for simulations and estimations) \cite{kuan1994implementing} propose a modification in the estimation method of the covariance matrix $\mathbf{\Omega}_{(n)}$. Moreover, \cite{chu1995moving} present the asymptotic properties of the ME process estimates and its limiting distribution (i.e., Brownian bridge increments). The ME detector is given by
\begin{align}
M_n (t|h)= \frac{\floor{nh}}{\hat{\sigma} \sqrt{n}} . \mathbf{\Omega}^{1/2}_{(n)} . \left[ \hat{\beta}_t^{(\floor{n t}-\floor{nh},\floor{nh} )} - \hat{\beta}_t^{(n)} \right] 
\end{align}
for some $h$, $0<h \leq 1$ the moving data window as a percentage of the historical period.  

The rescaled version of the covariance matrix $\mathbf{\Omega}_{(n)}$, which can adopted to both RE and ME processes and given by $\mathbf{\Omega}_{(\floor{nt} - \floor{nh}, \floor{nh} )}$ instead of using just the full historical period uses only the corresponding observations used in each moving iteration in order to rescale the estimate of the model parameter. Both covariances matrices converge asymptotically to the true covariance matrix, however \cite{kuan1994implementing} illustrate with a simulation study that the rescaled covariance matrix has faster rate of convergence and reduce bias, increasing the robustness and accuracy of the algorithm.     

\subsubsection{Sequential Residual-based processes}

The sequential residual-based processes are less computational expensive than the estimates-based processes since these processes monitor sequentially the residuals of a regression model computed at the beginning of the algorithm.    

\paragraph{OLS-CUSUM:} The OLS-CUSUM partial sum process consists of cumulative sums of standardized residuals and is often used in literature as a break point detector. In particular, \cite{brown1975techniques} study the properties of the (non-sequential) OLS-CUSUM and OLS-CUSUM squares. The detector function is given by the following expression 
\begin{align}
G_n (r) &= \frac{1}{\hat{\sigma} \sqrt{n}} . \sum_{t=1}^{\floor{nr}}\hat{\epsilon}_t^{(n)} , \ \text{where } \ \hat{\epsilon}_t^{(n)} = y_t - x_{t-1}^{\prime} \hat{\beta}_t^{(n)}, \ \ t=n+1,...,N.
\end{align}
Under the null hypothesis of no structural instability, the process $G_n (r)$ can be proved to converge into a Brownian bridge, $\sigma W^0 (t)$, which is the corresponding limiting process. Moreover, the OLS-CUSUM process is well suited to detect relatively short lasting structural instability and performs better for changes early in the monitoring period. 

Another important aspect of consideration, especially when monitoring time series regression models such as the stationary AR(1) model, is to provide a correction for the covariance matrix since the existence of a structural break can cause inflated values for $\hat{\rho}$ (close to unit root) and possibly contributing to a spurious unit root, which consequently can affect the power of the test statistic. In particular, \cite{andreou2008restoring} prove that a HAC estimator, using a boundary condition to control the divergence of the long term variance, can fix the monotonicity property of the power function and improve the asymptotic properties of the empirical size.  

\newpage 

Additionally, the (non-sequential) OLS-CUSUM of residual squares is given by 
\begin{align}
\Gamma \big( n, k \big) = \sum_{t=n+1}^{n+k} \left[  \hat{\epsilon}^2_t - \left( \frac{1}{n} \sum_{t=1}^n \hat{\epsilon}^2_t \right)^2 \right]
\end{align} 
with its limit distribution depending on the distribution of the errors and approximated by a Brownian bridge \cite{deng2008limit}. An appropriate modification of the above detector function along with an appropriate boundary function (e.g., $b_6(t))$ that has traceable crossing probabilities yields the corresponding sequential process suitable for the monitoring schemes we consider in this paper.     

\paragraph{OLS-MOSUM:} The OLS-MOSUM process is constructed based on sums of residuals, with the main difference between the OLS-CUSUM process to be the endogenous computation of the detector function, such that the least-square residuals are estimated using a regression model fitted to the stationary time series observations from each fixed bandwidth window $h$, where $0 < h \leq1$ across the monitoring period. 
\begin{align}
H_n (r| h) &= \frac{t}{\hat{\sigma} \sqrt{n}} \sum_{t=\floor{\xi r}-\floor{nh}+1}^{\floor{\xi r}} \hat{\epsilon}_t  ,\ \  \ \text{with } \ \ \hat{\epsilon}_t = y_t - x_{t-1}^{\prime} \hat{\beta}_t, \ \ \ \xi= \frac{n - \floor{nh}}{1-h}. 
\end{align}

Therefore, it can be easily proved that the OLS-MOSUM test statistic has a limiting process which corresponds to the Brownian bridge increments (see e.g., \cite{chu1995moving}). The known form of the limiting distribution simplifies statistical inference.  Furthermore, the OLS-MOSUM detector might be better suited for sequential detection of multiple breaks due to the nature of the algorithm. We can investigate the particular aspect in future research.  

\medskip

\begin{theorem}
\label{theorem 1}
Suppose that an FCLT and WLLN limit results hold, then it can be proved that the empirical fluctuation processes have the following limiting distributions
\begin{enumerate}
\item[\textit{(i)}] 
$R_n(r) \overset{d}{\to} W^{0}(r)$ for some $0 \leq r \leq 1-h$,

\item[\textit{(ii)}] 
$M_n (r|h) \overset{d}{\to} M(r|h)$, where $M(r|h) := W^{0}(r+h) - W^{0}(r)$ for $0 \leq r \leq 1-h$,   

\item[\textit{(iii)}]
$G_n (r) \overset{d}{\to} \sigma \ W^0 (r)$ for some $0 \leq r \leq 0$,

\item[\textit{(iv)}]
$H_n ( r| h ) = \displaystyle{ G_n \left( \frac{ \floor{\xi r }} {n} \right) - G_n \left( \frac{ \floor{\xi r } - \floor{ nr } } {n}  \right) \overset{d}{\to}} W^{0}(r+h) - W^{0}(r)$.
\end{enumerate}
\end{theorem}
Formal proofs for the limit results given by Theorem \ref{theorem 1} \textit{(i)} and \textit{(ii)} can be found in \cite{leisch2000monitoring}, while for (iii) and as extension for (iv) can be found in \cite{ploberger1992cusum}. In particular a sketch proof  for the limit result of Theorem \ref{theorem 1} \textit{(iii)} can be found in the Appendix of the paper. However, notice that a review of these limit results especially to accommodate correctly the features of these detectors within the monitoring period is essential to develop relevant asymptotic theory. We will examine the particular aspects in a future research paper.

\newpage 

\subsection{Crossing probabilities and boundary functions}
\label{boundaries}

Detecting structural breaks within the proposed framework implies the use of a sequential detection algorithm. The boundary functions are selected so that they satisfy certain regularity conditions such that the algorithm has a fast break point detection. Extensive examination of these regularity conditions are presented in the studies of \cite{robbins1970boundary}, \cite{lerche1986boundary} as well as in \cite{chu1996monitoring}. Appropriate conditions include the existence of a concave non-decreasing function $b(t)$ as $t \to \infty$ (see Lemma \ref{lemma:boundaries}). 

More precisely, under the null hypothesis of no parameter instability within the monitoring period, these boundary functions do not affect the empirical size of the test statistics since a pre-specified crossing probability (significance level) is defined to control the size. However, under the alternative hypothesis the power function of the test statistics is affected by the capability of the boundary to detect any early signs of structural instability in the parameters  of the model \citep{zeileis2005monitoring}. We consider the following linear and non-linear (parabolic and logarithmic) boundary functions for the empirical fluctuation processes.
\label{boundary 1}  
\begin{align}
b_1(t) = \sqrt{ t(t-1) \left[ \lambda^2 + \mathsf{log} \left( \frac{t}{t-1} \right) \right] } \ \ \text{ and } \ \ b_2(t) = \lambda \sqrt{ t(t-1)}
\end{align}
\begin{align}
b_3(t) = \lambda t, \ \ b_4(t) = \lambda t^2 \ \text{ and } \ b_5(t) = \lambda( t^2 - t + 0.1)
\end{align}
\begin{align}
b_6(t) = \lambda \sqrt{n} \left( 1 + \frac{t}{n} \right) \left( \frac{n}{t+n} \right)^\gamma
\ \text{where} \ t \geq n, \ \gamma \in (0,0.5] 
\end{align}
\begin{align}
b_7(t) = \lambda \left( t(t-0.5) \right)^{0.25} 
\end{align}
\begin{align}
b_8(t) = \lambda \big( t(t- \phi^-1) \big)^{\phi^{-1}} \ \text{ and } \ \ b_9(t) = \lambda \left( \frac{t}{t-\phi^-1} \right)^{\phi^{-1}}
\end{align}
where $\phi^{-1}=1.618$. We consider the growing rates of the boundary functions over time such that these correspond to monitoring periods with $T = 2$ and $T = 10$ as seen on Figure a and Figure b respectively. These are considered to be of order $\sqrt{t}$ and $\sqrt{\log(t)}$ respectively (see,  \citep{leisch2000monitoring}). Therefore, the specific boundary functions have different behaviour when employed as stopping rules within a sequential monitoring scheme, which depends on the location of the break point (such as early vis-a-vis late break), on the empirical process used (such as RE tends to be sensitive for late breaks) as well as the severity of the fluctuation (e.g., volatility persistence) and can be observed especially by the performance of the average run length as well as by empirical size or power distortions using a Monte Carlo simulation study. 

\newpage

\bigskip

\begin{lemma}
\label{lemma:boundaries}
Let $b(t)$ be a continuous positive defined boundary function used in a sequential empirical process, \cite{robbins1970boundary} proved that it has to satisfy the following conditions:
\begin{enumerate}
\item[\textit{(i)}] $\displaystyle t^{-1/2} b(t)$ is a non-decreasing function as $t \to \infty$,
\item[\textit{(ii)}] $\displaystyle t^{-1/2} b(t)$ is a non-increasing for t sufficiently small,
\item[\textit{(iii)}] $\displaystyle \int_{t_0}^{\infty} t^{-3/2} \  b(t) \exp \left\{ -\frac{1}{2t}  b^2(t) \right\}dt < \infty$, 
\item[\textit{(iv)}] $\displaystyle \int_{0}^{1} t^{-3/2} \  b(t) \exp \left\{ -\frac{1}{2t}  b^2(t) \right\}dt < \infty$.
\end{enumerate} 
\end{lemma}

Boundary $b_1(t)$ which has a closed form crossing probability is introduced by \cite{chu1996monitoring} and its applicability is also investigated by \cite{zeileis2005monitoring} which introduce the alternative boundary $b_3$ (no known closed form crossing probability). Furthermore, Boundary $b_2(t)$ is considered by \cite{zeileis2001p} and boundaries $b_4 (t)$ and $b_5 (t)$ are studied by \cite{zeileis2005unified}. The critical values for the boundaries $b_1 (t)$, $b_2 (t)$, $b_3 (t)$, $b_4 (t)$ and $b_5 (t)$  are $\lambda=$7.78, 3.15, 1.58, 2.49 and 6.043 respectively for significance level $\alpha = \%$. Boundary $b_6(t)$ is especially useful for sequential monitoring schemes and corresponding critical values for this boundary can be found on Table 1 in \cite{horvath2004monitoring} (e.g., for $\alpha=5\%$ and $\gamma=0.25$ then $\lambda=2.386$).  

Boundaries $b_7 (t)$, $b_8 (t)$ and $b_9 (t)$ are variations of the aforementioned boundaries and chosen to satisfy the properties of empirical size and power. Other boundary functions are examined by \cite{robbins1970boundary} include $b(t)= \sqrt{t[\lambda^2 + ln(t)]}$ and $b(t)= \sqrt{(t+1)[\lambda^2 + ln(t+1)]}$ as well as $b(t)= 2t \log(\log(t))$ by \cite{segen1980detecting}, which have a closed form, allowing for the computation of their critical values.
\begin{align}
\mathbb{P} \left\{|W(t)| \geq  \sqrt{t[\lambda^2 + ln(t)]}, t \geq 1 \right\}
&= 
2 \bigg[ 1 - \Phi(\alpha) + \alpha \ \phi(\alpha) \bigg]
\\
\mathbb{P} \left\{ |W(t)| \geq \sqrt{(t+1)[\lambda^2 + ln(t+1)]}, t \geq 0 \right\} 
&= 
\exp \bigg\{ \lambda^2/2 \bigg\}
\end{align} 
For sequentially monitoring financial indicators we mainly use boundary functions $b_1$ and $b_3$, although we consider other boundaries as well for simulation purposes. Table \ref{tab:b2 cv} gives the critical values of RE process for boundary $b_3$ for different monitoring periods $T$ and significance levels ($\alpha=5\%$ and $10\%$) simulated using 25,000 Brownian bridges\footnote{Table 1 provides simulation of critical values similar to the procedure proposed \cite{zeileis2005monitoring}. See also \cite{zeileis2001p} and the R package provided by  \cite{zeileis2002strucchange}.}.

\begin{table}[htbp]
  \centering
  \caption{Critical values for boundary $b_3$ }
    \begin{tabular}{|c|cccccccccc|}
    \hline
    \multirow{2}[0]{*}{$\alpha$} & \multicolumn{10}{c|}{T} \\
              & 1     & 2     & 3     & 4     & 5     & 6     & 7     & 8     & 9     & 10 \\
    \hline    
    5\%   & 0.062 & 1.577 & 1.835 & 1.940 & 2.005 & 2.039 & 2.070 & 2.097 & 2.111 & 2.128\\
    10\%  & 0.052 & 1.382 & 1.598 & 1.697 & 1.752 & 1.786 & 1.811 & 1.832 & 1.844 & 1.853 \\
    \hline
    \end{tabular}%
  \label{tab:b2 cv}%
\end{table}%

\newpage 

\section{Monte Carlo Simulation Study}
\label{Section3}

Our Monte Carlo Simulation Study aims to assess the statistical validity and effectiveness of the proposed break-point detection algorithms for finite (e.g., $N=200$) and asymptotic sample (e.g., $N=2000$) sizes via a simulated environment controlled by varying certain factors that represent stylized facts of financial data and in particular specific characteristics of financial indicators. These include the nature of the data generating process (DGP), (we illustrate and discuss the results for only one DGP for comparability purposes), the case of spurious breaks and existence of persistence in the model parameters, the break-point location, the sample size of historical and monitoring period (by varying the sampling frequency) which permits to obtain distribution of the detection delay of the empirical fluctuation processes.      

\subsection{Empirical size and power of testing hypothesis}

The computation of the empirical size, under the null hypothesis of structural stability, as well as of the power, under the alternative hypothesis of a single structural break at unknown break-point within the monitoring period is carried out with a Monte Carlo procedure. Statistically these can be considered as the probability of doing Type I Error and the complement of the probability of doing a Type II Error respectively, which can be helpful to control for algorithmic convergence using $B=2,500$ repetitions. Comparing the monotonicity property of the power function as well as checking for any empirical size distortions is a standard way to choose between competing empirical fluctuation processes \citep{stock1994unit}. 

More precisely, for the computation of the empirical size we keep the model parameters stable throughout the monitoring period while for the simulations of the power we introduce artificial breaks at pre-specified locations such that at 25$\%$, $50\%$ and $75\%$ of the monitoring period. Furthermore, we consider different sample sizes, such as $n=\{100,...,1000\}$, where for $N=200$ we represent a monthly frequency indicator for a period of about 16 years while for $N=1000$ we represent a weekly frequency indicator. Both computations can be thought in simple terms as percentages of the number of times the sequential processes detect structural change in the total times of algorithmic repetitions. We present the results of the simulations for the empirical size of the null hypothesis of no structural break, and the power of the alternative hypothesis of existence of an artificial structural break in the model, as these are defined by Assumption \ref{Assumption2} with $N=2n$. We also compute the detection of delay times of the empirical size and the power for the sequential processes using the boundary functions introduced in Section \ref{boundaries}. The data generating process (DGP) we use to carry out the simulations is given by
\begin{align}
y_t = \mu + \rho y_{t-1} + \epsilon_t, \ t=\{1,...,n,n+1,...,N \} \ \text{and} \ \ \epsilon_t \sim N(0,1) 
\end{align}
which produces normally distributed $\textit{i.i.d}$ random variables, that satisfy our assumptions. 

Under the null hypothesis, we set the model intercept with $\mu=1$ and the autocorrelation coefficient $\rho=\{0.3,0.5,0.7,0.9\}$ to consider the existence of different autocorrelation scenarios within the model such that for example, $\rho=0.3$ can be considered as a stationary model with $I(0)$ time series while $\rho=0.9$ can be considered as as a non-stationary model with parameter persistence, almost $I(1)$. Under the alternative hypothesis we consider two scenarios; (a) a structural break in the model intercept $\mu$ $(\mu=1 \to \mu=1.5)$; and (b) a structural break in the autocorrelation coefficient $\rho$ $(\rho=0.3 \to 0.5,0.6,0.7)$. 

An important point worth emphasizing when comparing between these different metrics for the sequential monitoring scheme is the fact that the suitability of a boundary function to detect structural changes depends on the detector function (due to the fact that these processes tend to have different type of fluctuations).  Furthermore, when examining the properties of empirical size we can verify whether the results are uniformly distributed, close to the nominal significance level $(\alpha=5\%)$ with low empirical size distortions, while checking also the performance the Average Run Length (ARL), that is, the break point detection timing for each process under each scenario. We obtain the simulated empirical size results for the OLS-CUSUM and RE test statistics using the boundary functions $b_1(t),b_2(t),b_3(t), b_7(t),b_8(t)$ and $b_9(t)$. Moreover, the empirical size results based on the quadratic boundary functions $b_4(t)$ and $b_5(t)$ have been also computed but since  they produce rather low empirical sizes we have not considered them further (seen also from their plots these boundaries level up too quickly missing out almost most of the fluctuations of the processes). Detailed describtion about the performance of boundary $b_6(t)$ in simulations is given by \cite{horvath2004monitoring}. We find that the boundary function $b_1(t)$ and $b_3(t)$ introduced by \cite{zeileis2005monitoring} provide good results, such that for the case of OLS-CUSUM and autocorrelation between 0.3 to 0.7 the empirical size is uniform especially after the finite sample size (i.e., $n=200)$.  

Under the alternative hypothesis, we obtain the power with sample size $n=\{50,100,200,1000\}$. These results are reported on the Tables found in the Appendix of the paper that correspond to structural breaks both in the model intercept and in the autocorrelation coefficients. Some  conclusions regarding the performance of power of the test statistics are: (a) increasing the sample size $n$, improves the detection performance significantly with initial values close to $10\%$ for small sample sizes to values close to 1 for large sample sizes, (b) the precision of power depends on the location of the structural change (for instance, structural break late in the monitoring period reduces predictive power) and (c) the value of the autocorrelation coefficient affects the accuracy of the algorithms (for example in the case of high autocorrelation coefficient such that $\rho=0.9$ and large sample sizes appears to induce some power distortions). Our simulation experiments verify that the power of the tests seem to satisfy the monotonicity property of the power functions, but with different rate of convergence across processes, boundary functions and break point location. The aforementioned conclusions can be also verified by observing the plots of power functions produced with $B=2,500$ replications for $n=\{100,...,1000\}$ (and $\rho=0.3)$. 

In particular, the linear boundary $b_3(t)$ works provides good empirical size results for both the OLS-CUSUM and RE test statistics for early and mid-way break-point locations (at $25\%$ and $50\%$ of the monitoring period). However, for the ME process and especially for late structural breaks the hyperbolic boundaries are superior to the linear boundary which also shows the robustness of the particular process for detecting changes with these boundaries late in the monitoring period. Similarly, for the OLS-MOSUM process, late changes within the monitoring period reduce the detection accuracy and robustness of the process with the linear boundary. Furthermore, the choice of the bandwidth window $h$ can affect the empirical power results. In particular, for the moving estimates processes, we set $h=0.5$, which is an optimal bandwidth choice that results to superior power when testing using the OLS-MOSUM test in comparison to the other test statistics such as the maximal likelihood ratio test (see,  \cite{chu1995moving}). Lastly, the asymptotic behaviour of the particular test statistic shows a high probability of detecting parameter instability within the monitoring period. 

\subsection{Average Run Length of empirical fluctuation processes}

The standard approach for assessing the performance of the detector functions in sequential analysis is to determine the distribution of the average run-length (ARL) within the monitoring period. Therefore, such a criterion to evaluate the proposed sequential processes is to test the effectiveness of the ARLs of the processes in detecting the break-point location. A relevant study in the literature is presented by  \cite{moustakides2004optimality}  who propose the min-max criterion for accessing the effectiveness of the CUSUM process. In this paper, we report the break-point detection percentages, the ARL distributions as well as their standard deviation for the null and alternative hypothesis on Tables \ref{tab:ARL1} and \ref{tab:ARL2} respectively. According to \cite{chu1996monitoring} a robust monitoring framework implies equivalent ARL distributions regardless of the break-point location (that is, asymptotic distributions which are similar to the ideal shape of the operating characteristic curve in an acceptance sampling scheme).  In our simulation study we consider the properties of ARLs for different break-point locations under the alternatives. 

Furthermore, we obtain the kernel densities of the simulated average delay times, which are useful for examining the attributes that affect the performance of ARL across the test statistics, as seen on Figure 1 and Figure 2. For example for   the OLS-CUSUM test with boundary $b_3$, the more negatively skewed is the location of the break in the monitoring period the lower the ARL appears to be. Although the process seems to have higher ARL for early changes within the monitoring period while induces an earlier indication for a later break, it seems to be of similar variability regardless of the location of the break (under $H_0$). Under the alternative as expected the kernel densities have heavier tails to the left due to the properties of the power functions. Our empirical findings as demonstrated by the kernel densities on the plots of Figure 1 and Figure 2 verify similar findings found in the literature (see, \cite{aue2009delay}). 

\medskip

Figure 1 (Kernel densities of delay times under null hypothesis). 

\medskip

Figure 2 (Kernel densities of delay times under alternative hypothesis). 

\medskip

Figure 3 (Power functions of empirical fluctuation processes with boundaries $b_1,b_2$ and $b_3$).

\bigskip

\subsection{Explanation of the monitoring procedure}

In this section we explain the procedure followed for monitoring the chosen indicators which allow us to shed some light regarding the dating and duration of the financial crisis of 2008 and the European sovereign debt crisis of 2010; while at the same time providing a way to evaluate the performance of the sequential break-point detection algorithms for detecting structural instability in the fitted models, such as the AR(1) model and the AR(1)-Garch(1,1) model.   

Model estimations and sequential monitoring were carried out between 1999:01 to 2016:05 while even though wider time intervals were considered such as 1985:01 to 2016:04 these results are not presented here since the focus of our work is on capturing the fluctuations of the financial crisis of 2008 and the spillover effects to the financial markets. However as pointed out by \cite{brave2012diagnosing} the particular sub-sampling approach contributes to a restricted information set by omitting past crises. Furthermore, by monitoring a sub-sample before the crisis, such as, 1999:01 to 2007:06 (or 1985:01 to 2007:06) we can verify the structural parameter instability with a comparison of the full sample model estimates and the corresponding sub-sample estimates as well as to check whether any other significant structural instabilities exist. 

\newpage

Therefore, by monitoring the aforementioned leading indicators we aim to justify their stochastic trends both statistically and economically and examine their behaviour ex ante and ex post. Summary tables with the model estimations and sequential monitoring results for both level and first difference series can be found in the Appendix of the paper.  

Firstly, we use the ADF unit root test, proposed by \cite{dickey1979distribution}, to check for the existence of unit roots in the time series observations. Our empirical findings show that for most of the indicators the null hypothesis of a unit root is not rejected, which provides statistical evidence of the permanent effect that the shock of the financial crisis had on these indicators as well as the fact that such a shock is not a realization of the underline data generating process but an exogeneous effect (see, \cite{perron1989great}). For example, the level series of the NFCI, STFSI and CISS indicators have ADF statistics -2.44, -2.82 and -2.08 respectively with p-values $> 0.10$ signalling the existence of a unit root. However, although the non-stationary level series appear to have high autocorrelations by taking first differences the stationarity assumption is ensured and satisfactory parameter estimates and HAC standard errors are obtained. Furthermore, we implement the breakpoint unit root test (with Schwarz IC) to identify the  structural break-point location for the trend stationary series either in 2007:06 (an early sign of structural instability of the indicators) or around the financial crisis.  

Secondly, the severity of the structural breaks of model parameters can be seen from the different values of the estimated model coefficients through a UD max test which computes the models between the regimes of the estimated breaks. More specifically, the particular testing procedure shows the number of breakpoints as determined by the unweighed maximized statistics. For example, implementing a UD max test with 3 breaks for the NFIC indicator the procedure estimates the corresponding model parameters of the 3 regimes (e.g., 2006:05 to 2008:11 is $y_t = 0.12 + 1.12y_{t-1}$ while for 2008:12 to 2011:12 is $y_t = -0.05 + 0.87y_{t-1}$) showing the changing stochastic trend of the specific indicator before and after the crisis. Furthermore, we implement retrospective sequential procedure proposed by \cite{bai1998estimating} which is suitable for testing for multiple breaks to identify the break-point location in the stationary time series of the indicators. The statistical significance of the breaks is also verified with information criteria, (IC), such as the Schwarz and LWZ\footnote{Notice that the IC are employed to estimate the number of breaks and include a term of the residual sum of squares for $s$ no. of breaks and a penalty term proportional to the number of parameters in the model. For example, \cite{hall2013inference} investigate the inferencing on structural breaks with the use of IC; such tests can particularly applied on interest rate series to test for structural instability in monetary policy.}. Our estimations illustrate that the LWZ statistic underestimates the number of breaks in the models, which is also verified by \cite{hall2013inference}. On the other hand, Schwarz is more robust in identifying a break within the duration interval of the financial crisis; which we define to be 2008:10 to 2009:03. However, notice that this is not a formal confidence interval but a common interval of the monitored indicators in which the structural instabilities appear to occur based on our empirical findings. 

Furthermore, we use an F-statistic as an extra retrospective testing method to test against a single-shift alternative of unknown timing  which also identifies a structural break within the crisis interval (e.g, for NFCI and its components a break location lies in the interval [2008:10,2008:12]. These particular F-statistics proposed by \cite{Andrews1993tests} compare residuals between a segmented regression and an unsegmented model (see also \cite{zeileis2005unified}). Additionally, as also discussed in \cite{andreou2009structural} multiple structural break testing encompasses two different procedures, the model selection based-approach and the sequential sample segmentation approach. Our study mainly focuses on the classification of retrospective vis-a-vis sequential monitoring procedures.

\newpage

Next, we implement the retrospective monitoring processes to the estimated models, which include the standard RE and OLS-CUSUM processes (with constant $b(t)=c_0$ and linear boundary $b(t)=\lambda t$) as well as the OLS-CUSUM of squared residuals. Our empirical results show that the OLS-CUSUM with both boundaries lack of any significant fluctuations, since the underline process is not crossing its $5\%$ level boundary, consequently a warning for structural instability is not issued (due to lack of power against shifts orthogonal to the mean regressor, which can be verified by computing the $\Delta \beta$ shift between estimated coefficients of historical and monitoring period (see, \cite{ploberger1990local}, \cite{zeileis2005monitoring}). On the other hand OLS-CUSUM-SQ has higher power (see \cite{deng2008limit}) in correctly detecting a break (e.g., for CISS EU index a break is detected on 2008:10). This conjecture is indeed verified by  \cite{ploberger1990local} who examine the local power of CUSUM and CUSUM-SQ showing that CUSUM-SQ has a limit distribution that depends on fourth moments of disturbances and thus a non-trivial power even against heteroscedasticity.

Thirdly, the next stage to retrospective monitoring is the implementation of the sequential monitoring procedure (based on the empirical fluctuation processes) to both the regression model (see Figure 4) and the residuals of Garch(1,1) of first difference series (see Figure 5). Specifically, we choose the monitoring period to be 2 times the historical period (i.e. $N=2n$) in order to match the monte carlo experiment of our simulation study. Thus, the period starting on 1999:01 up to 2004:10 is considered as the historical period while 2004:11 to 2016:05 the monitoring period; within which the values of the sequential processes are computed and monitored by checking whether they cross the pre-specified boundary function (using boundaries $b_1$ and $b_3$). For major indicators such as NFCI (and its components e.g., the risk index has similar behaviour as the main index), STFSI, Bofa, Moodys and CISS sequential break point detection is dated on 2008:10 (e.g., for monthly series, NFCI break is on 26/09/2008, also verified by \cite{brave2012diagnosing} as a day of financial turbulence). 

Furthermore, for less volatile indices such as the CFSI and KCFSI after first differencing, the sequential processes fail to detect the instability that is captured on their level series. Similarly, for some of the sub-components of the indicators for example the case of the financial market component of CISS, the fluctuations of the processes such as the OLS-CUSUM process seems to be on a lower level than the ones of the main indicators; which is not surprising since the sub-components have less features and cross-correlations and are less informative about the overall financial conditions and stability. As a result, the sequential processes fail to detect a true structural break; nevertheless is important to monitor also these indicators in order to verify our results and check which sub-series show financial stress.  

Our empirical study shows that monitoring the CISS Sovereign Stress Index component as well and corresponding indicators of main European economies permits to evaluate their reaction to economic events, such as the break-point location in comparison to the main index as well as the duration and severity of the crisis in regards to the Sovereign Debt crisis of 2010. Although large European economies such as Germany, had improvement in their financial conditions after the economic recession of 2008 (as seen from the level of the indicator), the Sovereign Debt stress crisis in Europe exhibits heterogeneity across European countries with different level of fluctuations and different break-point locations and severity (asymmetric effects).

\newpage

To conclude, our monitoring procedures are effective in capturing the signal of the financial crisis and match the events of September 2008 in US and are informative about the European Sovereign Debt crisis of 2010. Moreover, some of the conclusions we get from the sequential monitoring procedure include the following. Firstly, the choice of the boundary which affects the detection delay and type II error seems to depend on both the sampling frequency as well as the size of the historical period in comparison to the sample size of the monitoring period. For example a higher sampling frequency (such as weekly instead of monthly) seems to require larger historical period for empirical fluctuation procedures such as OLS-CUSUM to run effectively. Secondly, the choice of the process itself seems to depend on the nature of financial data under consideration. For example the OLS-CUSUM test statistic shows to be robust in detecting parameter instabilities with low delay especially for indicators such as NFCI and STFSI which show to have high volatility change and spike over a certain period of time. On the other hand, empirical fluctuation processes such as the OLS-MOSUM and ME are shown to be affected by the type of financial data as well as the position of the structural break within the monitoring period. Lastly, the choice of a dynamic boundary that moves along the trend of the data and depends on the movement of the series historical observations is an area worth investigating further. 

In addition, we monitor the model residuals fitted to level series produced from the fitted regression model as well as the residuals of the GARCH(1,1) model of the first difference series which appear to work well. In particular, for the non-linear time series we  estimate the models for monthly frequency data. However, for example for the CISS indicator we estimate the Garch(1,1) model for both weekly and monthly frequency in order to examine the effect of sampling frequency on the robustness of the model parameters as well as the power of the sequential detection algorithms. Estimation results can be found in the Appendix of the paper.

\medskip

Figure 4 (Different Sequential Monitoring scenarios of weekly and monthly NFCI level series).

\medskip

Figure 5 (Sequential Monitoring of Garch(1,1) residuals of main indicators).

\bigskip

\begin{small}
Figure 5 shows the monitoring of Garch(1,1) residuals of first difference monthly series. Time period for top 4 plots: 1999:01 to 2016:04. Time period for bottom 2 plots: 2000:09 to 2016:06. Break dates: (a) US indicators: STFSI 2008:09, NFCI, Moodys, Bofa 2008:10 (b) EU indicators: CISS Sovereign 2008:08, Germany CISS Sovereign 2007:09, with OLS-CUSUM and boundaries $b_1$, $b_3$.
\end{small}

\newpage

\section{Economic and Monetary Policy Discussion}
\label{Section4}

In this section we provide an overview of other important metrics which reflect economic conditions and financial stability. Firstly, monitoring financial conditions and stability indicators to provide a timely diagnosis of financial crisis is necessary for effective economic planning and regulation. Furthermore, monitoring both types of indicators simultaneously is essential in order to avoid ambiguous signals of a financial turbulence. In particular, economists argue regarding the correct set of economic policies which are required to dampen the effects of recessions and help economies to follow a sustainable growth path. 

\cite{romer1994ends} points out that opinions about fiscal policy range from  \textit{``the view that no recession has ever ended without fiscal expansion to the view that fiscal stimulus has always come too late".} Similarly, monetary policy and macroeconomic shocks can affect expectations of economic agents. Therefore, monitoring FCIs provides useful insights regarding the effectiveness of monetary policies such as the level of the its impact on economic activity and the improvement of economic conditions (see, \cite{hatzius2010financial}). In particular, for the case of US the low interest rates due to the US housing finance policy has been one of the contributing factors triggering the LB financial crisis and thus indicating the importance of designing flexible monetary policies and regulatory financial policies to minimize the risks of similar financial events in the future.     

Secondly, examining interest rates expectations in relation to forward rates is an additional method to view the severity of financial crisis through the lens of investors and consumers expectations about economic conditions as well as a tool to identify regimes of economic and monetary policies. According to \cite{friedman1979interest}, the empirical literature has tested various hypotheses about forward and expected future rates however no common consensus exists due to the difficulty in evaluating market participants' expectations. As a matter of fact, the author's empirical study show evidence of rejecting the pure expectations theory of the term structure of the interest rates. According to the author the systemic variation of the term structure is not independent of indicators of economic activity and monetary policy. The aforementioned studies provide us with evidence that sequentially monitoring such financial time series should reveal similar behaviour to the FSIs.     

Furthermore, as seen from our simulation study and the sequential monitoring of the ECIs and FSIs, the effectiveness or the negative impact of an economic policy can be seen from the severity and the duration of the structural change. A structural change that lasted for several months might be an indication of an economic policy (or monetary) not so antirecessionary. \cite{romer1994ends} finds evidence of large consistent declines on interest rates during recessions. Thus, to demonstrate this finding, we sequentially monitor two measures of inflation rates in the US, the 5 Year Breakeven Inflation Rate and the 5 Year Forward Inflation Expectations which permits to verify whether conclusions we obtain from the monitoring procedure are consistent with relevant macroeconomic theory about monetary policies and inflation expectations during periods of an economic crisis and prolonged uncertainty.

Our sequential break-point detection framework finds statistical evidence of a structural break within the crisis interval of 2008 for both rates. Specifically, for the 5 Year Breakeven Inflation Rate series, the structural break detected with the OLS-MOSUM (break due to changes in the level of the series) on the 26/09/2008 is the date on which the Washington Mutual Bank failure happened (largest failure in terms of assets, see e.g., \cite{brave2012diagnosing}).

\newpage

Therefore, our empirical findings verify the fact that during an economic crisis the inflation rates and inflation expectations are severely affected from available information about the severity of the crisis and thus exhibiting structural instability. In other words, people tend to form expectations with key macroeconomic relations; uncertainty about future inflation impose higher uncertainty about future interest rates. Such regimes of structural change in interest rates (i.e., severe decrease after the financial crisis) can be seen on Figure 6 (for Euro area, UK and US) and similarly to the EU government bond yields. While the former indices show the changes in monetary policy (in accordance also with the inflation expectations) the latter verify the deterioration of financial instability in Europe after the Sovereign Debt crisis of 2010 and their asymmetric effects across European economies. Similar asymmetric effects are found in government yields as described by \cite{giordano2012determinants}; such as evidence of time-dependent contagion component in sovereign spreads and different responses between core and peripheral countries. More specifically, this occurs due to mispricing of actual financial instability and revision of market expectations in regards to the severity of the crisis.      

In summary, since financial stability and monetary policy are two distinct types of regulatory and macro-prudential polices a holistic view in addition to just interest rate policies is required in order to understand possible contributing factors to an effective financial crisis management such as a more rational banking management and corporate responsibility. For example, \cite{michail2016lack} emphasize through their findings that monetary policy have no persistent impact on bank lending or bank credit risk taking thus lending decisions which affects the liquidity channel of banks should be explained otherwise. Another channel worth examining is the risk management optimization through Sovereign debt restructuring (see e.g., \cite{consiglio2015risk}\footnote{The authors use stochastic programming methodology to derive a model which accounts both the costs of debt financing and the associated risks aiming to choose a model for debt structure that has high probability of sustainable debt management.}). According to \cite{lane2012european}, in the Euro area level, new financial reforms are recommended and implemented  to better monitor the financial stability of the Euro area countries. Such policies include the monitor of a threshold for public debt levels, the use of common Eurobonds to manage over-borrowing by weaker economies across Europe as well as a dynamic collective financial policy across Euro area calibrated in relation to macroeconomic conditions. Furthermore, the cyclicality of the policies and the effect of austerity measures on helping economies to recover is a crucial dimension of the crisis management. In particular, \cite{vegh2014road} examine extensively how policy responses in Latin America have evolved over time (switching from procyclical to countercyclical). The author argues that the policymakers response in some Eurozone countries is of early Latin America type; showing that the implementation of procyclical fiscal policy in Eurozone had a negative economic impact by amplifying the duration and severity of the current crisis. As a matter of fact the aforementioned feedback loop is further fueled when there is inaction and uncertainty about the economic policy that regulatory authorities will follow (see, \cite{baker2016measuring}).


\newpage

\section{Conclusion}
\label{Section5}

An investigation of the symptoms causing financial and economic instability as well as unsustainable economic growth can be considered as the first stage of the problem identification. Emphasis should also be given to the next crucial stage, that is, the treatment and recovery stage. This stage includes the true identification of the factors and irregularities that can negatively contribute to the emergence of a financial and economic crisis. A careful examination of financial stress prevention and sustainable economic growth promotion  policies is necessary in order to prevent or mitigate the risk of crisis recurrence. Our study demonstrates some important empirical and simulation findings, verifying related studies in the literature.  
 
Firstly, the proposed sequential monitoring procedure such as the sequential monitoring of standardized residuals based on US and European indicators has shown to be particularly robust, in comparison to the timing of the financial crisis of 2008 in the US as well as the Sovereign Stress crisis of 2010 in Europe. The underling behaviour of these indicators which might cause any violations of the modelling assumptions does not appear to significantly compromise the predictive ability of the methodology, as also extensively examined with our simulation study. 

Secondly, the theoretical examination of the properties of sequential processes and their asymptotic distributions as well as our simulation study and the empirical comparisons of the sequential processes through the computation of the power functions and ARLs give us useful insights about the predictive performance of the detectors to identify structural instabilities in the time movement of indicators and especially during the economic crisis of 2008. Specifically, the robustness of the detectors depends on the size of the historical and monitoring periods as well as the location of the structural change within the monitoring period. For example, empirical fluctuation processes such as OLS-CUSUM and RE perform better for early changes in the monitoring period while the choice of the boundary function as well as the nature of financial data monitored (type of fluctuations, smoothness of transition between regimes, existence of volatility clustering) are further factors which can affect the performance of the tests.  

In particular, when the structural break was more severe with longer duration, then the tests could detect the break-point location earlier that it actually happens due to higher fluctuations of the induced monitoring values which do not much the curvature of the boundary function. Moreover, when these time series have a relative large conditional mean then a higher sampling frequency and a longer monitoring period will be required (such as Moodys and VIX financial series) to correctly identify the break-point location. On the other hand, in the case of quite low monitored values (e.g., when using standardized residuals of the Garch(1,1) model) a recursive estimates detector seems to perform better than other sequential processes. Thus, our empirical findings focus on sequentially monitoring regression residuals of level series as well as monitoring Garch(1,1) residuals of first difference series. 

Furthermore, the impact of sampling frequency is worth examining with respect to the accuracy of the empirical fluctuation processes in detecting structural breaks. Although we use monthly sampling frequency for the model estimation and monitoring of the indicators, an estimation comparison of available monthly frequency series yields indeed better statistical significance of some of the model estimates, reduces standard errors and in the sequential monitoring cases it give us more precise information regarding the dating of the break. Nevertheless, our study has some limitations as well, which we should point out. The accuracy of the sequential procedure depends on a correct model specification; the existence of model misspecification can affect the performance and robustness of the processes.


\newpage

In our study we use a relative simple linear model and for the case of non-linear models we consider the cases of Garch(1,1) and AR(1)-Garch(1,1); other stochastic volatility models for example could be considered, to further examine structural instability in the volatility. Last but not least, the uniqueness of these indicators lies on the fact that even though different construction methodologies are used in order to model the economic and financial conditions of highly heterogeneous markets their stochastic trend behaviour capture well the crisis events and are highly informative regarding for example existence of volatility instability. Further  limitations of our study include the fact that we fit the linear and non-linear time series regression models to first differenced time series observations. The particular practise implies that the possible presence of nonstationarity can not be correctly modelled. We leave the aspect of sequential monitoring under the presence of nonstationarity for future research.  
    
In comparison to the already existing methodologies in the literature our proposed testing framework is robust in terms of detection accuracy, parsimonious enough with no indication of the presence of multicollinearity. For example the choice of an optimal lag as well as the individual investigation of each indices ensures the avoidance of such econometric issues (see,  \cite{oet2013safe}). Adding a dynamic component of to an EWM substantially improves the quality of the results, such as it reduces false alarm ratio and increases the percentages of correctly predicting the structural break. Our proposed framework has interesting extensions supported with other novel econometric methods which can provide a suitable method for capturing dynamic and time-varying patterns of macroeconomic and behavioural variables.    

Given the global consequences of financial crises, is crucial for interconnected economies to collaborate and adopt common cross-border policies that can assess quickly and in a consistent manner any early signs of financial turbulence and thus take appropriate measures for avoiding the transmission of systemic risks between financial systems. Moreover, examining the impact of these structural instabilities of economic and financial conditions in relation to other economic metrics (e.g., interest rates and inflation expectations) is essential in order to implement appropriate financial and monetary policies to ensure stability of the system. 

Some further research worth mentioning include: (i) the endogenous estimation of conditional volatility within-monitoring
period and sequential extraction of standardized residuals, (ii) the modelling of structural instability in continuous-time models (such as stochastic volatility models), and (iii) the monitoring out-of sample economic conditions robust under the presence of structural change in mean and conditional variance. An appropriate sequential process along with a dynamic boundary function with a proven strong detecting ability can be chosen to monitor financial conditions using out-of sample data (real-time monitoring) providing this way an early warning sign for upcoming financial turbulence, so that appropriate correcting measures and economic policy is designed and implemented in a timely manner.

\begin{small}
\paragraph{Acknowledgements}

I would like to thank Prof. Jose Olmo and Prof. Jean-Yves Pitarakis from the Department of Economics, University of Southampton for helpful conversations. Furthermore, I wish to thank Prof. Elena Andreou  from the Department of Economics, University of Cyprus for guidance during the academic year 2015-2016. Also, I wish to thank Prof. Achim Zeileis from the Department of Statistics, University of Innsbruck for helpful discussion on the functionalities of the R package \texttt{strucchange}. This manuscript was compiled while being a PhD Candidate. Financial support from the VC PhD Scholarship of the University of Southampton is gratefully acknowledged. Financial support while working as a Research Assistant at the Economics Research Centre (UCY) from the ERC grant "MONITOR" of Prof. Elena Andreou is also gratefully acknowledged. 
\end{small}


\newpage

\bibliographystyle{apalike}

{\small
\bibliography{myreferences1}}

\begin{thebibliography}{}

\bibitem[Andreou, 2008]{andreou2008restoring}
Andreou, E. (2008).
\newblock Restoring monotone power in the cusum test.
\newblock {\em Economics Letters}, 98(1):48--58.

\bibitem[Andreou and Ghysels, 2006]{andreou2006monitoring}
Andreou, E. and Ghysels, E. (2006).
\newblock Monitoring disruptions in financial markets.
\newblock {\em Journal of Econometrics}, 135(1-2):77--124.

\bibitem[Andreou and Ghysels, 2009]{andreou2009structural}
Andreou, E. and Ghysels, E. (2009).
\newblock Structural breaks in financial time series.
\newblock {\em Handbook of financial time series}, pages 839--870.

\bibitem[Andrews, 1991]{andrews1991heteroskedasticity}
Andrews, D.~W. (1991).
\newblock Heteroskedasticity and autocorrelation consistent covariance matrix
  estimation.
\newblock {\em Econometrica: Journal of the Econometric Society}, pages
  817--858.

\bibitem[Andrews, 1993]{Andrews1993tests}
Andrews, D.~W. (1993).
\newblock Tests for parameter instability and structural change with unknown
  change point.
\newblock {\em Econometrica}, pages 821--856.

\bibitem[Aue and Horv{\'a}th, 2013]{aue2013structural}
Aue, A. and Horv{\'a}th, L. (2013).
\newblock Structural breaks in time series.
\newblock {\em Journal of Time Series Analysis}, 34(1):1--16.

\bibitem[Aue et~al., 2009]{aue2009delay}
Aue, A., Horv{\'a}th, L., and Reimherr, M.~L. (2009).
\newblock Delay times of sequential procedures for multiple time series
  regression models.
\newblock {\em Journal of Econometrics}, 149(2):174--190.

\bibitem[Babeck{\`y} et~al., 2012]{babecky2012banking}
Babeck{\`y}, J., Havranek, T., Mateju, J., Rusn{\'a}k, M., Smidkova, K., and
  Vasicek, B. (2012).
\newblock Banking, debt and currency crises: early warning indicators for
  developed countries.

\bibitem[Bai, 1997]{bai1997estimating}
Bai, J. (1997).
\newblock Estimating multiple breaks one at a time.
\newblock {\em Econometric theory}, 13(3):315--352.

\bibitem[Bai and Perron, 1998]{bai1998estimating}
Bai, J. and Perron, P. (1998).
\newblock Estimating and testing linear models with multiple structural
  changes.
\newblock {\em Econometrica}, pages 47--78.

\bibitem[Baker et~al., 2016]{baker2016measuring}
Baker, S.~R., Bloom, N., and Davis, S.~J. (2016).
\newblock Measuring economic policy uncertainty.
\newblock {\em The quarterly journal of economics}, 131(4):1593--1636.

\bibitem[Berkes et~al., 2004]{berkes2004sequential}
Berkes, I., Gombay, E., Horv{\'a}th, L., and Kokoszka, P. (2004).
\newblock Sequential change-point detection in garch (p, q) models.
\newblock {\em Econometric theory}, 20(6):1140--1167.

\bibitem[Brave and Butters, 2012]{brave2012diagnosing}
Brave, S. and Butters, R.~A. (2012).
\newblock Diagnosing the financial system: financial conditions and financial
  stress.
\newblock {\em International Journal of Central Banking}.

\bibitem[Brave and Butters, 2011]{brave2011monitoring}
Brave, S.~A. and Butters, R.~A. (2011).
\newblock Monitoring financial stability: A financial conditions index
  approach.
\newblock {\em Economic Perspectives}, 35(1):22.

\bibitem[Brodsky and Darkhovsky, 2013]{brodsky2013nonparametric}
Brodsky, E. and Darkhovsky, B.~S. (2013).
\newblock {\em Nonparametric methods in change point problems}, volume 243.
\newblock Springer Science \& Business Media.

\bibitem[Brown et~al., 1975]{brown1975techniques}
Brown, R.~L., Durbin, J., and Evans, J.~M. (1975).
\newblock Techniques for testing the constancy of regression relationships over
  time.
\newblock {\em Journal of the Royal Statistical Society: Series B
  (Methodological)}, 37(2):149--163.

\bibitem[Chen, 2002]{chen2002tests}
Chen, M.-y. (2002).
\newblock Tests for parameter constancy in regression models.
\newblock {\em Department of Economics, National Chung Cheng University}.

\bibitem[Chen et~al., 2010]{chen2010sequentialmon}
Chen, Z., Tian, Z., Qin, R., et~al. (2010).
\newblock Sequential monitoring variance change in linear regression model.
\newblock In {\em J. Math. Res. Exposition}, volume~30, pages 610--618.

\bibitem[Chochola, 2008]{chochola2008sequential}
Chochola, O. (2008).
\newblock Sequential monitoring for change in scale.
\newblock {\em Kybernetika}, 44(5):715--730.

\bibitem[Chow, 1960]{chow1960tests}
Chow, G.~C. (1960).
\newblock Tests of equality between sets of coefficients in two linear
  regressions.
\newblock {\em Econometrica: Journal of the Econometric Society}, pages
  591--605.

\bibitem[Chu et~al., 1995a]{chu1995mosum}
Chu, C.-S.~J., Hornik, K., and Kaun, C.-M. (1995a).
\newblock Mosum tests for parameter constancy.
\newblock {\em Biometrika}, 82(3):603--617.

\bibitem[Chu et~al., 1995b]{chu1995moving}
Chu, C.-S.~J., Hornik, K., and Kuan, C.-M. (1995b).
\newblock The moving-estimates test for parameter stability.
\newblock {\em Econometric Theory}, 11(4):699--720.

\bibitem[Chu et~al., 1996]{chu1996monitoring}
Chu, C.-S.~J., Stinchcombe, M., and White, H. (1996).
\newblock Monitoring structural change.
\newblock {\em Econometrica: Journal of the Econometric Society}, pages
  1045--1065.

\bibitem[Cogley and Nason, 1995]{cogley1995output}
Cogley, T. and Nason, J.~M. (1995).
\newblock Output dynamics in real-business-cycle models.
\newblock {\em The American Economic Review}, pages 492--511.

\bibitem[Consiglio and Zenios, 2015]{consiglio2015risk}
Consiglio, A. and Zenios, S.~A. (2015).
\newblock Risk management optimization for sovereign debt restructuring.
\newblock {\em Journal of Globalization and Development}, 6(2):181--213.

\bibitem[Cs{\"o}rg{\H{o}} et~al., 2003]{csorgHo2003donsker}
Cs{\"o}rg{\H{o}}, M., Szyszkowicz, B., Wu, Q., et~al. (2003).
\newblock Donsker's theorem for self-normalized partial sums processes.
\newblock {\em The Annals of Probability}, 31(3):1228--1240.

\bibitem[Deng and Perron, 2008]{deng2008limit}
Deng, A. and Perron, P. (2008).
\newblock The limit distribution of the cusum of squares test under general
  mixing conditions.
\newblock {\em Econometric Theory}, 24(3):809--822.

\bibitem[Dickey and Fuller, 1979]{dickey1979distribution}
Dickey, D.~A. and Fuller, W.~A. (1979).
\newblock Distribution of the estimators for autoregressive time series with a
  unit root.
\newblock {\em Journal of the American statistical association},
  74(366a):427--431.

\bibitem[Doz et~al., 2012]{doz2012quasi}
Doz, C., Giannone, D., and Reichlin, L. (2012).
\newblock A quasi--maximum likelihood approach for large, approximate dynamic
  factor models.
\newblock {\em Review of economics and statistics}, 94(4):1014--1024.

\bibitem[Dufour, 1982]{dufour1982generalized}
Dufour, J.-M. (1982).
\newblock Generalized chow tests for structural change: A coordinate-free
  approach.
\newblock {\em International Economic Review}, pages 565--575.

\bibitem[Estrella and Mishkin, 1998]{estrella1998predicting}
Estrella, A. and Mishkin, F.~S. (1998).
\newblock Predicting us recessions: Financial variables as leading indicators.
\newblock {\em Review of Economics and Statistics}, 80(1):45--61.

\bibitem[Frankel and Saravelos, 2012]{frankel2012can}
Frankel, J. and Saravelos, G. (2012).
\newblock Can leading indicators assess country vulnerability? evidence from
  the 2008--09 global financial crisis.
\newblock {\em Journal of International Economics}, 87(2):216--231.

\bibitem[Friedman, 1979]{friedman1979interest}
Friedman, B.~M. (1979).
\newblock Interest rate expectations versus forward rates: Evidence from an
  expectations survey.
\newblock {\em The Journal of Finance}, 34(4):965--973.

\bibitem[Giordano et~al., 2012]{giordano2012determinants}
Giordano, L., Linciano, N., and Soccorso, P. (2012).
\newblock The determinants of government yield spreads in the euro area.

\bibitem[Gramlich and Oet, 2011]{gramlich2011structural}
Gramlich, D. and Oet, M.~V. (2011).
\newblock The structural fragility of financial systems: Analysis and modeling
  implications for early warning systems.
\newblock {\em The Journal of Risk Finance}, 12(4):270--290.

\bibitem[Hall et~al., 2013]{hall2013inference}
Hall, A.~R., Osborn, D.~R., and Sakkas, N. (2013).
\newblock Inference on structural breaks using information criteria.
\newblock {\em The Manchester School}, 81:54--81.

\bibitem[Hanschel et~al., 2005]{hanschel2005measuring}
Hanschel, E., Monnin, P., et~al. (2005).
\newblock Measuring and forecasting stress in the banking sector: evidence from
  switzerland.
\newblock {\em BIS papers}, 22:431--449.

\bibitem[Hansen, 2000]{hansen2000testing}
Hansen, B.~E. (2000).
\newblock Testing for structural change in conditional models.
\newblock {\em Journal of Econometrics}, 97(1):93--115.

\bibitem[Hansen, 2001]{hansen2001new}
Hansen, B.~E. (2001).
\newblock The new econometrics of structural change: Dating breaks in us labor
  productivity.
\newblock {\em The Journal of Economic Perspectives}, 15(4):117--128.

\bibitem[Hatzius et~al., 2010]{hatzius2010financial}
Hatzius, J., Hooper, P., Mishkin, F.~S., Schoenholtz, K.~L., and Watson, M.~W.
  (2010).
\newblock Financial conditions indexes: A fresh look after the financial
  crisis.
\newblock Technical report, National Bureau of Economic Research.

\bibitem[Hollo et~al., 2012]{hollo2012ciss}
Hollo, D., Kremer, M., and Lo~Duca, M. (2012).
\newblock Ciss-a composite indicator of systemic stress in the financial
  system.

\bibitem[Horv{\'a}th et~al., 2004]{horvath2004monitoring}
Horv{\'a}th, L., Hu{\v{s}}kov{\'a}, M., Kokoszka, P., and Steinebach, J.
  (2004).
\newblock Monitoring changes in linear models.
\newblock {\em Journal of Statistical Planning and Inference}, 126(1):225--251.

\bibitem[Horv{\'a}th et~al., 2001]{horvath2001empirical}
Horv{\'a}th, L., Teyssi{\`e}re, G., et~al. (2001).
\newblock Empirical process of the squared residuals of an arch sequence.
\newblock {\em The Annals of Statistics}, 29(2):445--469.

\bibitem[Hu{\v{s}}kov{\'a} and Chochola, 2010]{huvskova2010simple}
Hu{\v{s}}kov{\'a}, M. and Chochola, O. (2010).
\newblock Simple sequential procedures for change in distribution.
\newblock In {\em Nonparametrics and Robustness in Modern Statistical Inference
  and Time Series Analysis: A Festschrift in honor of Professor Jana
  Jure{\v{c}}kov{\'a}}, pages 95--104. Institute of Mathematical Statistics.

\bibitem[Koopman et~al., 2012]{koopman2012dynamic}
Koopman, S.~J., Lucas, A., and Schwaab, B. (2012).
\newblock Dynamic factor models with macro, frailty, and industry effects for
  us default counts: the credit crisis of 2008.
\newblock {\em Journal of Business \& Economic Statistics}, 30(4):521--532.

\bibitem[Kr{\"a}mer et~al., 1988]{kramer1988testing}
Kr{\"a}mer, W., Ploberger, W., and Alt, R. (1988).
\newblock Testing for structural change in dynamic models.
\newblock {\em Econometrica: Journal of the Econometric Society}, pages
  1355--1369.

\bibitem[Kr{\"a}mer et~al., 1991]{kramer1991recursive}
Kr{\"a}mer, W., Ploberger, W., and Schl{\"u}ter, I. (1991).
\newblock Recursive vs. ols residuals in the cusum test.
\newblock In {\em Economic Structural Change}, pages 35--47. Springer.

\bibitem[Kuan and Chen, 1994]{kuan1994implementing}
Kuan, C.-M. and Chen, M.-Y. (1994).
\newblock Implementing the fluctuation and moving-estimates tests in dynamic
  econometric models.
\newblock {\em Economics Letters}, 44(3):235--239.

\bibitem[Kulperger et~al., 2005]{kulperger2005high}
Kulperger, R., Yu, H., et~al. (2005).
\newblock High moment partial sum processes of residuals in garch models and
  their applications.
\newblock {\em The Annals of Statistics}, 33(5):2395--2422.

\bibitem[Kushner and Yin, 2003]{kushner2003stochastic}
Kushner, H. and Yin, G.~G. (2003).
\newblock {\em Stochastic approximation and recursive algorithms and
  applications}, volume~35.
\newblock Springer Science \& Business Media.

\bibitem[Lane, 2012]{lane2012european}
Lane, P.~R. (2012).
\newblock The european sovereign debt crisis.
\newblock {\em Journal of economic perspectives}, 26(3):49--68.

\bibitem[Leisch et~al., 2000]{leisch2000monitoring}
Leisch, F., Hornik, K., and Kuan, C.-M. (2000).
\newblock Monitoring structural changes with the generalized fluctuation test.
\newblock {\em Econometric Theory}, 16(6):835--854.

\bibitem[Lerche, 1986]{lerche1986boundary}
Lerche, H.~R. (1986).
\newblock {\em Boundary crossing of Brownian motion: Its relation to the law of
  the iterated logarithm and to sequential analysis}, volume~40.
\newblock Springer Science \& Business Media.

\bibitem[Michail et~al., 2016]{michail2016lack}
Michail, N.~A., Koursaros, D., Savva, C.~S., et~al. (2016).
\newblock The lack of persistence of interest rate changes on banks lending and
  risk taking behaviour.
\newblock {\em Bulletin of Economic Research}, 12273.

\bibitem[Moustakides, 2004]{moustakides2004optimality}
Moustakides, G.~V. (2004).
\newblock Optimality of the cusum procedure in continuous time.
\newblock {\em The Annals of Statistics}, 32(1):302--315.

\bibitem[Newey and West, 1986]{newey1986simple}
Newey, W.~K. and West, K.~D. (1986).
\newblock A simple, positive semi-definite, heteroskedasticity and
  autocorrelationconsistent covariance matrix.

\bibitem[Oet et~al., 2013]{oet2013safe}
Oet, M.~V., Bianco, T., Gramlich, D., and Ong, S.~J. (2013).
\newblock Safe: An early warning system for systemic banking risk.
\newblock {\em Journal of Banking \& Finance}, 37(11):4510--4533.

\bibitem[Perron, 1989]{perron1989great}
Perron, P. (1989).
\newblock The great crash, the oil price shock, and the unit root hypothesis.
\newblock {\em Econometrica: journal of the Econometric Society}, pages
  1361--1401.

\bibitem[Pitarakis, 2004]{pitarakis2004least}
Pitarakis, J.-Y. (2004).
\newblock Least squares estimation and tests of breaks in mean and variance
  under misspecification.
\newblock {\em The Econometrics Journal}, 7(1):32--54.

\bibitem[Ploberger and Kr{\"a}mer, 1990]{ploberger1990local}
Ploberger, W. and Kr{\"a}mer, W. (1990).
\newblock The local power of the cusum and cusum of squares tests.
\newblock {\em Econometric Theory}, 6(3):335--347.

\bibitem[Ploberger and Kr{\"a}mer, 1992]{ploberger1992cusum}
Ploberger, W. and Kr{\"a}mer, W. (1992).
\newblock The cusum test with ols residuals.
\newblock {\em Econometrica: Journal of the Econometric Society}, pages
  271--285.

\bibitem[Robbins and Siegmund, 1970]{robbins1970boundary}
Robbins, H. and Siegmund, D. (1970).
\newblock Boundary crossing probabilities for the wiener process and sample
  sums.
\newblock {\em The Annals of Mathematical Statistics}, pages 1410--1429.

\bibitem[Romer and Romer, 1994]{romer1994ends}
Romer, C.~D. and Romer, D.~H. (1994).
\newblock What ends recessions?
\newblock {\em NBER macroeconomics annual}, 9:13--57.

\bibitem[Segen and Sanderson, 1980]{segen1980detecting}
Segen, J. and Sanderson, A. (1980).
\newblock Detecting change in a time-series (corresp.).
\newblock {\em IEEE Transactions on Information Theory}, 26(2):249--254.

\bibitem[Siegmund, 1986]{siegmund1986boundary}
Siegmund, D. (1986).
\newblock Boundary crossing probabilities and statistical applications.
\newblock {\em The Annals of Statistics}, pages 361--404.

\bibitem[Stock, 1994]{stock1994unit}
Stock, J.~H. (1994).
\newblock Unit roots, structural breaks and trends.
\newblock {\em Handbook of econometrics}, 4:2739--2841.

\bibitem[Stock and Watson, 1989]{stock1989new}
Stock, J.~H. and Watson, M.~W. (1989).
\newblock New indexes of coincident and leading economic indicators.
\newblock {\em NBER macroeconomics annual}, 4:351--394.

\bibitem[Vegh and Vuletin, 2014]{vegh2014road}
Vegh, C.~A. and Vuletin, G. (2014).
\newblock The road to redemption: Policy response to crises in latin america.
\newblock {\em IMF Economic Review}, 62(4):526--568.

\bibitem[Watson and Engle, 1983]{watson1983alternative}
Watson, M.~W. and Engle, R.~F. (1983).
\newblock Alternative algorithms for the estimation of dynamic factor, mimic
  and varying coefficient regression models.
\newblock {\em Journal of Econometrics}, 23(3):385--400.

\bibitem[Watson et~al., 1991]{watson1991using}
Watson, M.~W. et~al. (1991).
\newblock Using econometric models to predict recessions.
\newblock {\em Economic Perspectives}, 15(Nov):14--25.

\bibitem[Zeileis, 2001]{zeileis2001p}
Zeileis, A. (2001).
\newblock p values and alternative boundaries for cusum tests.
\newblock Technical report, Technical Report.

\bibitem[Zeileis, 2005]{zeileis2005unified}
Zeileis, A. (2005).
\newblock A unified approach to structural change tests based on ml scores, f
  statistics, and ols residuals.
\newblock {\em Econometric Reviews}, 24(4):445--466.

\bibitem[Zeileis et~al., 2002]{zeileis2002strucchange}
Zeileis, A., Leisch, F., Hornik, K., and Kleiber, C. (2002).
\newblock strucchange: An r package for testing for structural change in linear
  regression models.
\newblock {\em Journal of statistical software}, 7(1):1--38.

\bibitem[Zeileis et~al., 2005]{zeileis2005monitoring}
Zeileis, A., Leisch, F., Kleiber, C., and Hornik, K. (2005).
\newblock Monitoring structural change in dynamic econometric models.
\newblock {\em Journal of Applied Econometrics}, 20(1):99--121.

\bibitem[Zeileis et~al., 2010]{zeileis2010testing}
Zeileis, A., Shah, A., and Patnaik, I. (2010).
\newblock Testing, monitoring, and dating structural changes in exchange rate
  regimes.
\newblock {\em Computational Statistics \& Data Analysis}, 54(6):1696--1706.

\end{thebibliography}

\newpage

\section{Appendix}

The Appendix of the paper presents main simulation, estimation and asymptotic theory results relevant to our econometric analysis. In particular, Appendix \ref{AppendixA} presents main simulations results for the Monte Carlo simulation study of the paper such as power simulations and simulated average run length of the detection of the break-point under the alternative hypothesis. Appendix \ref{AppendixB} presents the main estimation results of our empirical study which considers the sequential monitoring of economic and financial indicators before and after the 2008 financial crisis in US and Europe. Appendix \ref{AppendixD} provides background asymptotic theory results and some useful asymptotic results.

\paragraph{Simulations Results:} 

All simulations results such as the empirical size under the null hypothesis and the power under the alternative hypothesis as well as ARLs distributions under the null and under the alternative hypothesis are carried out using the R package \texttt{strucchange} with $B=2,500$ Monte Carlo replications.   

Consider the model
\begin{align}
y_t = \mu + \rho y_{t-1} + \epsilon_t, \epsilon_t \sim N(0,1)
\end{align}

with \textit{i.i.d} normally distributed errors, which produce a normally distributed response variable with and mean different than zero. Furthermore, tor the empirical fluctuation processes, we use a a significance level $\alpha=5\%$, with monitoring period 2 times the historical period $(N=2n)$, a moving window $h=0.5$ (for the ME and OLS-MOSUM) and rescaling for the estimates-based processes (for the ME and RE test statistics).

\paragraph{Estimation Results:} 

Statistical significance of the estimates is denoted with the order: ``***'' if p-value $\leq 0.01$,``**'' if p-value $\leq 0.05$  and ``*'' if p-value $\leq 0.1$. 

The covariance matrix method employed for the computation of the standard errors of the regression estimates is the HAC\footnotemark (Newey-West) of fixed 12 lags (for the monthly frequency indicators). Similarly the parameter estimates of the Garch(1,1) and AR(1)-Garch(1,1) regression models with normally distributed innovations we report the robust standard errors, which are calculated using the R package \texttt{rugarch}.

For unit root testing an ADF test is used as well as a unit root break point test with a regression model with trend and intercept (in Eviews) and similar model for the retrospective F-test. For the sequential (retrospective) detection of multiple breaks tests we used the Global Information Criterion with maximum number of breaks to be 3 and a trimming percentage to be $15\%$. For the standard process monitoring we report the OLS-CUSUM-SQ (in Eviews) and for the sequential monitoring processes we report our estimates from our algorithm in R.

\footnotetext{In particular, the studies of \cite{newey1986simple} and \cite{andrews1991heteroskedasticity} propose a framework for the covariance matrix construction of parameter estimators in linear and non-linear regression models in the presence of heteroscedastisity and autocorrelation as well as the bandwidth estimators used to control convergence and increase estimation accuracy.} 

\appendix

\newpage

\begin{landscape}

\section{Simulation Results}
\label{AppendixA}

\subsection{Power functions and Average Run Length}

\begin{small}

\begin{table}[h!]
  \centering
  \caption{Power functions under the alternative hypothesis with break of $\rho$ } 
    \begin{tabular}{|c|ccccccccccccccccc|}
     \multicolumn{1}{c}{\textbf{}} & \multicolumn{1}{c}{\textbf{}} &       & \multicolumn{7}{c}{\textbf{Location 25\%}}            &       & \multicolumn{7}{c}{\textbf{Location 50\%}} \\
    \hline
          &       &       &       &       &       &       &       &       &       &       &       &       &       &       &       &       &  \\
          &       &       & \multicolumn{3}{c}{\textbf{OLS-CUSUM}} &       & \multicolumn{3}{c}{\textbf{RE}} &       & \multicolumn{3}{c}{\textbf{OLS-CUSUM}} &       & \multicolumn{3}{c|}{\textbf{RE}} \\
         \hline
          &       &       & \multicolumn{3}{c}{\textbf{Break in autocorrelation }} &       & \multicolumn{3}{c}{\textbf{Break in autocorrelation }} &       & \multicolumn{3}{c}{\textbf{Break in autocorrelation }} &       & \multicolumn{3}{c|}{\textbf{Break in autocorrelation }} \\
    \multicolumn{1}{|c|}{\textbf{b(t)}} & \multicolumn{1}{c}{$\textbf{n}$} &       & \multicolumn{1}{c}{\textbf{0.5}} & \multicolumn{1}{c}{\textbf{0.6}} & \multicolumn{1}{c}{\textbf{0.7}} &       & \multicolumn{1}{c}{\textbf{0.5}} & \multicolumn{1}{c}{\textbf{0.6}} & \multicolumn{1}{c}{\textbf{0.7}} &       & \multicolumn{1}{c}{\textbf{0.5}} & \multicolumn{1}{c}{\textbf{0.6}} & \multicolumn{1}{c}{\textbf{0.7}} &       & \multicolumn{1}{c}{\textbf{0.5}} & \multicolumn{1}{c}{\textbf{0.6}} & \multicolumn{1}{c|}{\textbf{0.7}} \\
    \hline
    \multicolumn{1}{|c|}{\multirow{4}[8]{*}{\textbf{1}}} & \multicolumn{1}{c}{\textbf{50}} &       & \multicolumn{1}{c}{0.1840} & \multicolumn{1}{c}{0.4328} & \multicolumn{1}{c}{0.7956} &       & \multicolumn{1}{c}{0.2592} & \multicolumn{1}{c}{0.5308} & \multicolumn{1}{c}{0.8568} &       & \multicolumn{1}{c}{0.0936} & \multicolumn{1}{c}{0.2176} & \multicolumn{1}{c}{0.5020} &       & \multicolumn{1}{c}{0.1680} & \multicolumn{1}{c}{0.3252} & \multicolumn{1}{c|}{0.6584} \\
    \multicolumn{1}{|c|}{} & \multicolumn{1}{c}{\textbf{100}} &       & \multicolumn{1}{c}{0.2604} & \multicolumn{1}{c}{0.6648} & \multicolumn{1}{c}{0.9696} &       & \multicolumn{1}{c}{0.3772} & \multicolumn{1}{c}{0.8056} & \multicolumn{1}{c}{0.9912} &       & \multicolumn{1}{c}{0.1212} & \multicolumn{1}{c}{0.3480} & \multicolumn{1}{c}{0.7752} &       & \multicolumn{1}{c}{0.2236} & \multicolumn{1}{c}{0.5484} & \multicolumn{1}{c|}{0.9128} \\
    \multicolumn{1}{|c|}{} & \multicolumn{1}{c}{\textbf{200}} &       & \multicolumn{1}{c}{0.4392} & \multicolumn{1}{c}{0.9160} & \multicolumn{1}{c}{1.0000} &       & \multicolumn{1}{c}{0.6220} & \multicolumn{1}{c}{0.9720} & \multicolumn{1}{c}{0.9996} &       & \multicolumn{1}{c}{0.1756} & \multicolumn{1}{c}{0.5880} & \multicolumn{1}{c}{0.9620} &       & \multicolumn{1}{c}{0.3588} & \multicolumn{1}{c}{0.8124} & \multicolumn{1}{c|}{0.9960} \\
    \multicolumn{1}{|c|}{} & \multicolumn{1}{c}{\textbf{1000}} &       & \multicolumn{1}{c}{0.9916} & \multicolumn{1}{c}{1.0000} & \multicolumn{1}{c}{1.0000} &       & \multicolumn{1}{c}{1.0000} & \multicolumn{1}{c}{1.0000} & \multicolumn{1}{c}{1.0000} &       & \multicolumn{1}{c}{0.7612} & \multicolumn{1}{c}{1.0000} & \multicolumn{1}{c}{1.0000} &       & \multicolumn{1}{c}{0.9664} & \multicolumn{1}{c}{1.0000} & \multicolumn{1}{c|}{1.0000} \\
    \hline
    \multicolumn{1}{|c|}{\multirow{4}[8]{*}{\textbf{2}}} & \multicolumn{1}{c}{\textbf{50}} &       & \multicolumn{1}{c}{0.1092} & \multicolumn{1}{c}{0.3124} & \multicolumn{1}{c}{0.7116} &       & \multicolumn{1}{c}{0.1800} & \multicolumn{1}{c}{0.4132} & \multicolumn{1}{c}{0.7824} &       & \multicolumn{1}{c}{0.0544} & \multicolumn{1}{c}{0.1308} & \multicolumn{1}{c}{0.3936} &       & \multicolumn{1}{c}{0.1184} & \multicolumn{1}{c}{0.2352} & \multicolumn{1}{c|}{0.5444} \\
    \multicolumn{1}{|c|}{} & \multicolumn{1}{c}{\textbf{100}} &       & \multicolumn{1}{c}{0.1620} & \multicolumn{1}{c}{0.5412} & \multicolumn{1}{c}{0.9416} &       & \multicolumn{1}{c}{0.2668} & \multicolumn{1}{c}{0.6972} & \multicolumn{1}{c}{0.9760} &       & \multicolumn{1}{c}{0.0736} & \multicolumn{1}{c}{0.2348} & \multicolumn{1}{c}{0.6600} &       & \multicolumn{1}{c}{0.1576} & \multicolumn{1}{c}{0.4304} & \multicolumn{1}{c|}{0.8576} \\
    \multicolumn{1}{|c|}{} & \multicolumn{1}{c}{\textbf{200}} &       & \multicolumn{1}{c}{0.3016} & \multicolumn{1}{c}{0.8460} & \multicolumn{1}{c}{0.9976} &       & \multicolumn{1}{c}{0.4916} & \multicolumn{1}{c}{0.9456} & \multicolumn{1}{c}{0.9996} &       & \multicolumn{1}{c}{0.0912} & \multicolumn{1}{c}{0.4472} & \multicolumn{1}{c}{0.9344} &       & \multicolumn{1}{c}{0.2640} & \multicolumn{1}{c}{0.7072} & \multicolumn{1}{c|}{0.9916} \\
    \multicolumn{1}{|c|}{} & \multicolumn{1}{c}{\textbf{1000}} &       & \multicolumn{1}{c}{0.9724} & \multicolumn{1}{c}{1.0000} & \multicolumn{1}{c}{1.0000} &       & \multicolumn{1}{c}{0.9984} & \multicolumn{1}{c}{1.0000} & \multicolumn{1}{c}{1.0000} &       & \multicolumn{1}{c}{0.6048} & \multicolumn{1}{c}{0.9992} & \multicolumn{1}{c}{1.0000} &       & \multicolumn{1}{c}{0.9136} & \multicolumn{1}{c}{1.0000} & \multicolumn{1}{c|}{1.0000} \\
    \hline
    \multicolumn{1}{|c|}{\multirow{4}[8]{*}{\textbf{3}}} & \multicolumn{1}{c}{\textbf{50}} &       & \multicolumn{1}{c}{0.2940} & \multicolumn{1}{c}{0.5928} & \multicolumn{1}{c}{0.8860} &       & \multicolumn{1}{c}{0.3572} & \multicolumn{1}{c}{0.6620} & \multicolumn{1}{c}{0.9264} &       & \multicolumn{1}{c}{0.1852} & \multicolumn{1}{c}{0.3568} & \multicolumn{1}{c}{0.6800} &       & \multicolumn{1}{c}{0.2348} & \multicolumn{1}{c}{0.4588} & \multicolumn{1}{c|}{0.7952} \\
    \multicolumn{1}{|c|}{} & \multicolumn{1}{c}{\textbf{100}} &       & \multicolumn{1}{c}{0.4156} & \multicolumn{1}{c}{0.8190} & \multicolumn{1}{c}{0.9868} &       & \multicolumn{1}{c}{0.5220} & \multicolumn{1}{c}{0.9016} & \multicolumn{1}{c}{0.9968} &       & \multicolumn{1}{c}{0.2324} & \multicolumn{1}{c}{0.5300} & \multicolumn{1}{c}{0.8884} &       & \multicolumn{1}{c}{0.3140} & \multicolumn{1}{c}{0.7124} & \multicolumn{1}{c|}{0.9664} \\
    \multicolumn{1}{|c|}{} & \multicolumn{1}{c}{\textbf{200}} &       & \multicolumn{1}{c}{0.6324} & \multicolumn{1}{c}{0.9636} & \multicolumn{1}{c}{1.0000} &       & \multicolumn{1}{c}{0.7780} & \multicolumn{1}{c}{0.9940} & \multicolumn{1}{c}{1.0000} &       & \multicolumn{1}{c}{0.3684} & \multicolumn{1}{c}{0.7612} & \multicolumn{1}{c}{0.9880} &       & \multicolumn{1}{c}{0.5320} & \multicolumn{1}{c}{0.9200} & \multicolumn{1}{c|}{0.9992} \\
    \multicolumn{1}{|c|}{} & \multicolumn{1}{c}{\textbf{1000}} &       & \multicolumn{1}{c}{0.9972} & \multicolumn{1}{c}{1.0000} & \multicolumn{1}{c}{1.0000} &       & \multicolumn{1}{c}{1.0000} & \multicolumn{1}{c}{1.0000} & \multicolumn{1}{c}{1.0000} &       & \multicolumn{1}{c}{0.9016} & \multicolumn{1}{c}{1.0000} & \multicolumn{1}{c}{1.0000} &       & \multicolumn{1}{c}{0.9940} & \multicolumn{1}{c}{1.0000} & \multicolumn{1}{c|}{1.0000} \\
    \hline
    \end{tabular}%
  \label{tab:power5}%
\end{table}%

\underline{Notes:} Model under alternative hypothesis is $y_t = 1 + 0.3 y_{t-1} + \epsilon_t, \epsilon_t \sim N(0,1)$ with break of $\rho$ (0.3 $\to$ 0.5, 0.3 $\to$ 0.6 and  0.3 $\to$ 0.7) applied with $\alpha=5\%$ and $B=2500$ repetitions.

\end{small}

\end{landscape}


\newpage

\begin{landscape}

\begin{small}

\begin{table}[h!]
  \centering
  \caption{Average Run Length under the null hypothesis}
\renewcommand\tabcolsep{3pt}      
    \begin{tabular}{|r|rrcccrcccrcccrcccrcccrccc|}
    \hline
          &       &       & \multicolumn{7}{c}{\textbf{Location 25\%}}            &       & \multicolumn{7}{c}{\textbf{Location 50\%}}            &       & \multicolumn{7}{c|}{\textbf{Location 75\%}} \\
    \hline
          &       &       & \multicolumn{3}{c}{\textbf{OLS-CUSUM}} & \textbf{} & \multicolumn{3}{c}{\textbf{RE}} &       & \multicolumn{3}{c}{\textbf{OLS-CUSUM}} & \textbf{} & \multicolumn{3}{c}{\textbf{RE}} &       & \multicolumn{3}{c}{\textbf{OLS-CUSUM}} & \textbf{} & \multicolumn{3}{c|}{\textbf{RE}} \\
          \hline
    \multicolumn{1}{|c|}{\textbf{b(t)}} & \multicolumn{1}{c}{\textbf{n}} &       & \textbf{\%} & \textbf{ARL} & \textbf{s.d} & \textbf{} & \textbf{\%} & \textbf{ARL} & \textbf{s.d} &       & \textbf{\%} & \textbf{ARL} & \textbf{s.d} & \textbf{} & \textbf{\%} & \textbf{ARL} & \textbf{s.d} &       & \textbf{\%} & \textbf{ARL} & \textbf{s.d} & \textbf{} & \textbf{\%} & \textbf{ARL} & \textbf{s.d} \\
    \hline
    \multicolumn{1}{|c|}{\multirow{4}[8]{*}{\textbf{1}}} & \multicolumn{1}{c}{\textbf{50}} &       & 4     & 4     & 13    &       & 10    & -4    & 10    &       & 4     & -8    & 13    &       & 10    & -17   & 10    &       & 4     & -21   & 13    &       & 10    & -29   & 10 \\
    \multicolumn{1}{|c|}{} & \multicolumn{1}{c}{\textbf{100}} &       & 4     & 10    & 28    &       & 10    & -10   & 21    &       & 4     & -15   & 28    &       & 10    & -35   & 21    &       & 4     & -40   & 28    &       & 10    & -60   & 21 \\
    \multicolumn{1}{|c|}{} & \multicolumn{1}{c}{\textbf{200}} &       & 3     & 11    & 52    &       & 11    & -16   & 48    &       & 3     & -39   & 52    &       & 11    & -66   & 48    &       & 3     & -89   & 52    &       & 11    & -116  & 48 \\
    \multicolumn{1}{|c|}{} & \multicolumn{1}{c}{\textbf{1000}} &       & 3     & 30    & 267   &       & 11    & -108  & 231   &       & 3     & -220  & 267   &       & 11    & -358  & 231   &       & 3     & -470  & 267   &       & 11    & -608  & 231 \\
    \hline
    \multicolumn{1}{|c|}{\multirow{4}[8]{*}{\textbf{2}}} & \multicolumn{1}{c}{\textbf{50}} &       & 2     & 0     & 12    &       & 8     & -7    & 7     &       & 2     & -12   & 12    &       & 8     & -19   & 7     &       & 2     & -25   & 12    &       & 8     & -32   & 7 \\
    \multicolumn{1}{|c|}{} & \multicolumn{1}{c}{\textbf{100}} &       & 2     & 3     & 27    &       & 8     & -15   & 18    &       & 2     & -22   & 27    &       & 8     & -40   & 18    &       & 2     & -47   & 27    &       & 9     & -65   & 18 \\
    \multicolumn{1}{|c|}{} & \multicolumn{1}{c}{\textbf{200}} &       & 2     & -9    & 45    &       & 9     & -34   & 29    &       & 2     & -59   & 45    &       & 9     & -84   & 29    &       & 2     & -109  & 45    &       & 9     & -134  & 29 \\
    \multicolumn{1}{|c|}{} & \multicolumn{1}{c}{\textbf{1000}} &       & 2     & -63   & 230   &       & 10    & -198  & 136   &       & 2     & -313  & 230   &       & 10    & -448  & 136   &       & 2     & -563  & 230   &       & 11    & -698  & 136 \\
    \hline
    \multicolumn{1}{|c|}{\multirow{4}[8]{*}{\textbf{3}}} & \multicolumn{1}{c}{\textbf{50}} &       & 7     & 19    & 11    &       & 8     & 15    & 13    &       & 7     & 7     & 11    &       & 8     & 3     & 13    &       & 7     & -6    & 11    &       & 8     & -10   & 13 \\
    \multicolumn{1}{|c|}{} & \multicolumn{1}{c}{\textbf{100}} &       & 6     & 36    & 23    &       & 10    & 36    & 25    &       & 6     & 11    & 23    &       & 10    & 11    & 25    &       & 6     & -14   & 23    &       & 10    & -14   & 25 \\
    \multicolumn{1}{|c|}{} & \multicolumn{1}{c}{\textbf{200}} &       & 5     & 79    & 41    &       & 10    & 79    & 46    &       & 5     & 29    & 41    &       & 10    & 29    & 46    &       & 5     & -21   & 41    &       & 10    & -21   & 46 \\
    \multicolumn{1}{|c|}{} & \multicolumn{1}{c}{\textbf{1000}} &       & 5     & 415   & 212   &       & 10    & 409   & 200   &       & 5     & 165   & 212   &       & 10    & 159   & 200   &       & 5     & -85   & 212   &       & 10    & -91   & 200 \\
    \hline
    \end{tabular}%
  \label{tab:ARL1}%
\end{table}%
The model under the alternative hypothesis is $y_t = \mu + 0.3 \ y_{t-1} + \epsilon_t, \epsilon_t \sim N(0,1)$ with break of $\mu$ (1 $\to$ 1.5), with a significance level $\alpha=5\%$ for the OLS-CUSUM test and $\alpha=10\%$ for the RE test and $B=2,500$ Monte Carlo replications. Percentages presented on the table are rounded to the nearest integers.

\end{small}

\end{landscape}


\newpage

\begin{landscape}

\begin{small}

\begin{table}[h!]
  \centering
  \caption{Average Run Length under the alternative hypothesis with break of $\mu$ (1 $\to$ 1.5)}
\renewcommand\tabcolsep{3pt}   
    \begin{tabular}{|r|rrcccrcccrcccrcccrcccrccc|}
    \hline
          &       &       & \multicolumn{7}{c}{\textbf{Location 25\%}}            &       & \multicolumn{7}{c}{\textbf{Location 50\%}}            &       & \multicolumn{7}{c|}{\textbf{Location 75\%}} \\
   \hline
          &       &       & \multicolumn{3}{c}{\textbf{OLS-CUSUM}} & \textbf{} & \multicolumn{3}{c}{\textbf{RE}} &       & \multicolumn{3}{c}{\textbf{OLS-CUSUM}} & \textbf{} & \multicolumn{3}{c}{\textbf{RE}} &       & \multicolumn{3}{c}{\textbf{OLS-CUSUM}} & \textbf{} & \multicolumn{3}{c|}{\textbf{RE}} \\
          \hline
    \multicolumn{1}{|c|}{\textbf{b(t)}} & \multicolumn{1}{c}{\textbf{n}} &       & \textbf{\%} & \textbf{ARL} & \textbf{s.d} & \textbf{} & \textbf{\%} & \textbf{ARL} & \textbf{s.d} &       & \textbf{\%} & \textbf{ARL} & \textbf{s.d} & \textbf{} & \textbf{\%} & \textbf{ARL} & \textbf{s.d} &       & \textbf{\%} & \textbf{ARL} & \textbf{s.d} & \textbf{} & \textbf{\%} & \textbf{ARL} & \textbf{s.d} \\
    \hline
    \multicolumn{1}{|c|}{\multirow{4}[8]{*}{\textbf{1}}} & \multicolumn{1}{c}{\textbf{50}} &       & 27    & 18    & 12    & \multicolumn{1}{c}{} & 24    & 8     & 14    &       & 11    & 9     & 14    &       & 14    & -5    & 17    &       & 5     & -13   & 17    &       & 10    & -23   & 16 \\
    \multicolumn{1}{|c|}{} & \multicolumn{1}{c}{\textbf{100}} &       & 48    & 39    & 21    & \multicolumn{1}{c}{} & 42    & 23    & 29    &       & 20    & 24    & 26    &       & 24    & -3    & 36    &       & 6     & -15   & 36    &       & 14    & -48   & 34 \\
    \multicolumn{1}{|c|}{} & \multicolumn{1}{c}{\textbf{200}} &       & 82    & 73    & 38    & \multicolumn{1}{c}{} & 71    & 59    & 51    &       & 39    & 57    & 39    &       & 39    & 27    & 66    &       & 9     & -10   & 67    &       & 16    & -65   & 82 \\
    \multicolumn{1}{|c|}{} & \multicolumn{1}{c}{\textbf{1000}} &       & 100   & 134   & 74    & \multicolumn{1}{c}{} & 100   & 105   & 125   &       & 100   & 196   & 119   &       & 99    & 138   & 218   &       & 48    & 137   & 174   &       & 57    & 14    & 336 \\
    \hline
    \multicolumn{1}{|c|}{\multirow{4}[8]{*}{\textbf{2}}} & \multicolumn{1}{c}{\textbf{50}} &       & 17    & 19    & 12    & \multicolumn{1}{c}{} & 17    & 6     & 14    &       & 6     & 8     & 15    &       & 11    & -8    & 17    &       & 3     & -16   & 18    &       & 8     & -29   & 12 \\
    \multicolumn{1}{|c|}{} & \multicolumn{1}{c}{\textbf{100}} &       & 32    & 42    & 21    & \multicolumn{1}{c}{} & 28    & 19    & 31    &       & 11    & 21    & 29    &       & 16    & -11   & 37    &       & 2     & -26   & 39    &       & 10    & -55   & 31 \\
    \multicolumn{1}{|c|}{} & \multicolumn{1}{c}{\textbf{200}} &       & 69    & 82    & 38    & \multicolumn{1}{c}{} & 53    & 58    & 56    &       & 24    & 60    & 42    &       & 28    & 6     & 76    &       & 4     & -27   & 80    &       & 13    & -98   & 74 \\
    \multicolumn{1}{|c|}{} & \multicolumn{1}{c}{\textbf{1000}} &       & 100   & 159   & 79    & \multicolumn{1}{c}{} & 100   & 124   & 146   &       & 99    & 233   & 122   &       & 97    & 148   & 257   &       & 31    & 142   & 201   &       & 40    & -87   & 414 \\
    \hline
    \multicolumn{1}{|c|}{\multirow{4}[8]{*}{\textbf{3}}} & \multicolumn{1}{c}{\textbf{50}} &       & 46    & 21    & 9     & \multicolumn{1}{c}{} & 39    & 19    & 11    &       & 23    & 14    & 8     &       & 25    & 10    & 10    &       & 11    & 0     & 10    &       & 14    & -2    & 12 \\
    \multicolumn{1}{|c|}{} & \multicolumn{1}{c}{\textbf{100}} &       & 71    & 41    & 18    & \multicolumn{1}{c}{} & 65    & 39    & 20    &       & 39    & 29    & 16    &       & 37    & 24    & 18    &       & 13    & 7     & 18    &       & 18    & 1     & 22 \\
    \multicolumn{1}{|c|}{} & \multicolumn{1}{c}{\textbf{200}} &       & 93    & 68    & 31    & \multicolumn{1}{c}{} & 90    & 68    & 35    &       & 65    & 59    & 27    &       & 64    & 52    & 31    &       & 22    & 17    & 35    &       & 28    & 13    & 39 \\
    \multicolumn{1}{|c|}{} & \multicolumn{1}{c}{\textbf{1000}} &       & 100   & 139   & 53    & \multicolumn{1}{c}{} & 100   & 135   & 62    &       & 100   & 167   & 78    &       & 100   & 149   & 89    &       & 73    & 140   & 91    &       & 77    & 119   & 116 \\
    \hline
    \end{tabular}%
  \label{tab:ARL2}%
\end{table}%

The model under the alternative hypothesis is $y_t = \mu + 0.3 \ y_{t-1} + \epsilon_t, \epsilon_t \sim N(0,1)$ with break of $\mu$ (1 $\to$ 1.5), with a significance level $\alpha=5\%$ for the OLS-CUSUM test and $\alpha=10\%$ for the RE test and $B=2,500$ Monte Carlo replications. Percentages presented on the table are rounded to the nearest integers.

\end{small}

\end{landscape}


\newpage

\section{Estimation Results}
\label{AppendixB}

\subsection{Summary Statistics}

\bigskip

\begin{table}[h!]
  \centering
  \caption{Descriptive Statistics for  Monthly Average frequency Indicators}
  \renewcommand\tabcolsep{3pt}
    \begin{tabular}{|c|cccccccc|}
    \hline
    \textbf{Series} & \textbf{N} & \textbf{mean } & \textbf{Trim mean} & \textbf{St dev} & \textbf{Min} & \textbf{Max} & \textbf{Skewness } & \textbf{Kurtosis} \\
    \hline
    \textbf{NFCI} & 208   & -0.339 & -0.4221 & 0.613 & -0.96 & 2.77  & 2.78  & 9.43 \\
    \textbf{RISK} & 208   & -0.362 & -0.4432 & 0.599 & -0.99 & 2.75  & 2.79  & 9.73 \\
    \textbf{CREDIT} & 208   & -0.246 & -0.3244 & 0.604 & -0.89 & 2.64  & 2.46  & 7.63 \\
    \textbf{LEVERAGE} & 208   & -0.071 & -0.125 & 0.919 & -1.8  & 3.71  & 1.37  & 3.33 \\
    \hline
    \textbf{STFSI} & 208   & -0.12 & -0.2172 & 1.098 & -1.611 & 5.096 & 1.64  & 4.53 \\
    \textbf{CFSI} & 208   & 0.287 & 0.2706 & 0.939 & -1.92 & 2.89  & 0.33  & -0.38 \\
    \textbf{KCFSI} & 208   & 0.178 & 0.0239 & 1.134 & -1.07 & 6.15  & 2.64  & 9.07 \\
    \hline
    \textbf{BOFA} & 209   & 1.554 & 1.3997 & 1.035 & 0.67  & 6.78  & 2.86  & 9.73 \\
    \textbf{TED} & 209   & 0.464 & 0.4058 & 0.423 & 0.12  & 3.35  & 2.98  & 12.55 \\
    \textbf{MOODYS} & 209   & 2.687 & 2.6159 & 0.772 & 1.55  & 6.01  & 1.57  & 4.56 \\
    \textbf{VIX} & 209   & 20.84 & 20.022 & 8.295 & 10.82 & 62.64 & 1.93  & 5.89 \\
    \hline
    \textbf{CISS EU} & 209   & 0.198 & 0.1795 & 0.168 & 0.033 & 0.778 & 1.6   & 2.02 \\
    \textbf{CISS MONEY} & 209   & 0.044 & 0.04218 & 0.025 & 0.013 & 0.14  & 1.4   & 2.18 \\
    \textbf{CISS FINANCE} & 209   & 0.114 & 0.11105 & 0.063 & 0.023 & 0.285 & 0.63  & -0.5 \\
    \textbf{CISS BOND} & 209   & 0.04  & 0.03922 & 0.023 & 0.008 & 0.1   & 0.49  & -0.71 \\
    \hline
    \end{tabular}%
  \label{tab:addlabel}%
\end{table}%

\begin{small}
\underline{Notes:} Due to the existence of extreme values (especially during the period of economic crises of 2008) the indicators are positively skewed with high kurtosis (in many cases more than 3, the kurtosis for normally distributed data). The calculation of the trimmed mean (i.e., the mean after excluding the $10\%$ extreme tails of the distribution of the indicators) signals the severity of the shocks to the expected mean value of the financial indicators, which is also shown by the high deviation of their maximum values from the mean. 
\end{small}

\newpage

\begin{landscape}

\subsection{Monitoring Economic and Financial Indicators}

\begin{table}[h!]
  \centering
  \caption{US Financial Conditions Indicator NFCI level series and its components}
  \renewcommand\tabcolsep{3pt}
\begin{tabular}{cccccccccccccc}
    \hline
    \multicolumn{2}{c}{\textbf{Financial Series}} & \multicolumn{3}{c}{\textbf{NFCI}} & \multicolumn{3}{c}{\textbf{NFCI Risk}} & \multicolumn{3}{c}{\textbf{NFCI Credit}} & \multicolumn{3}{c}{\textbf{NFCI Leverage}} \\
    \hline
    \multicolumn{2}{c}{\textbf{Unit root testing}} & \textbf{ADF} & \textbf{p-value} & \textbf{Date} & \textbf{ADF} & \textbf{p-value} & \textbf{Date} & \textbf{ADF} & \textbf{p-value} & \textbf{Date} & \textbf{ADF} & \textbf{p-value} & \textbf{Date} \\
    \hline
    \multicolumn{2}{c}{\textbf{Unit root test}} & -2.4353 & 0.3603 & -     & -2.3092 &  0.4266 & -     & -2.6303 &  0.2674 &       & -2.7444 & 0.2201 & - \\
    \multicolumn{2}{c}{\textbf{Break point UR test}} & -5.0919 & 0.0632* & Jun-07 & -4.9359 & 0.0895* & Jul-07 & -4.5937 & 0.1957 & Jun-07 & -4.0863 & 0.4644 & May-07 \\
    \multicolumn{14}{c}{\textbf{}} \\
    \hline
    \multicolumn{2}{c}{\textbf{Model Estimation }} & \textbf{est} & \textbf{HAC s.e} & \textbf{p-value} & \textbf{est} & \textbf{HAC s.e} & \textbf{p-value} & \textbf{est} & \textbf{HAC s.e} & \textbf{p-value} & \textbf{est} & \textbf{HAC s.e} & \textbf{p-value} \\
    \hline
    \multirow{2}[2]{*}{Regression model } & $\mu$ & -0.0128 & 0.0238 & 0.5901 & -0.0155 & 0.0232 & 0.5053 & -0.0065 & 0.0186 & 0.7269 & -0.0084 & 0.0218 & 0.6987 \\
          & $\rho$ & 0.9682 & 0.0302 & *** & 0.9647 & 0.0280 & *** & 0.9744 & 0.0264 & *** & 0.9755 & 0.0253 & *** \\
\hline
    \end{tabular}%
  \label{tab:addlabel}%
\end{table}%
\end{landscape}


\newpage

\begin{landscape}
\begin{table}[h!]
  \centering
  \caption{US Financial Stability Indicators level series (St.Louis, Clevland, Kansas City)}
   \renewcommand\tabcolsep{3pt}
    \begin{tabular}{ccccccccccc}
    \hline
    \multicolumn{2}{c}{\textbf{Financial Series}} & \multicolumn{3}{c}{\textbf{STLFSI}} & \multicolumn{3}{c}{\textbf{CFSI}} & \multicolumn{3}{c}{\textbf{KCFSI}} \\
    \hline
    \multicolumn{2}{c}{\textbf{Unit root testing}} & \textbf{ADF} & \textbf{p-value} & \textbf{Date} & \textbf{ADF} & \textbf{p-value} & \textbf{Date} & \textbf{ADF} & \textbf{p-value} & \textbf{Date} \\
    \hline
    \multicolumn{2}{c}{\textbf{Unit root test}} & -2.8186 & 0.1923 & -     & -3.2224 &  0.0829* & -     & -3.1915 &  0.0890* & - \\
    \multicolumn{2}{c}{\textbf{Break point UR test}} & -4.8187 & <0.05* & Oct-08 & -4.5245 &  0.2266 & Jun-07 & -4.7452 & 0.1418 & Jun-07 \\
    \multicolumn{11}{c}{\textbf{}} \\
    \hline
    \multicolumn{2}{c}{\textbf{Model Estimation }} & \textbf{est} & \textbf{HAC s.e} & \textbf{p-value} & \textbf{est} & \textbf{HAC s.e} & \textbf{p-value} & \textbf{est} & \textbf{HAC s.e} & \textbf{p-value} \\
    \hline
    \multirow{2}[2]{*}{Regression model } & $\mu$ & -0.0106 & 0.0245 & 0.6668 & 0.0325 & 0.0319 & 0.3089 & 0.0045 & 0.0275 & 0.8690 \\
          & $\rho$ & 0.9605 & 0.0192 & ***   & 0.8973 & 0.0283 & ***   & 0.9495 & 0.0260 & *** \\
\hline
    \end{tabular}%
  \label{tab:addlabel}%
\end{table}%
\end{landscape}


\newpage

\begin{landscape}

\begin{table}[htbp]
  \centering
  \caption{US Financial Spread, yield and Volatility Indicators level series}
  \renewcommand\tabcolsep{3pt}
    \begin{tabular}{cccccccccccccc}
    \hline
    \multicolumn{2}{c}{\multirow{4}[2]{*}{\textbf{Financial Series}}} & \multicolumn{3}{c}{\textbf{BofA Merrill Lynch US }} &       &       &       & \multicolumn{3}{c}{\textbf{}} &       &       &  \\
    \multicolumn{2}{c}{} & \multicolumn{3}{c}{\textbf{Corporate 3-5 Year }} & \multicolumn{3}{c}{\textbf{Ted Spread}} & \multicolumn{3}{c}{\textbf{Moody's Seasoned Baa }} & \multicolumn{3}{c}{\textbf{CBOE Volatility Index}} \\
    \multicolumn{2}{c}{} & \multicolumn{3}{c}{\textbf{Option-Adjusted Spread}} & \multicolumn{3}{c}{\textbf{}} & \multicolumn{3}{c}{\textbf{Corporate Bond Yield }} & \multicolumn{3}{c}{\textbf{}} \\
    \hline
    \multicolumn{2}{c}{\textbf{Unit root testing}} & \textbf{ADF} & \textbf{p-value} & \textbf{Date} & \textbf{ADF} & \textbf{p-value} & \textbf{Date} & \textbf{ADF} & \textbf{p-value} & \textbf{Date} & \textbf{ADF} & \textbf{p-value} & \textbf{Date} \\
    \hline
    \multicolumn{2}{c}{\textbf{Unit root test}} & -2.9692 & 0.1435 & -     & -2.9528 & 0.1484 & -     & -3.2782 &  0.0728* & -     & -3.5043 & ** & - \\
    \multicolumn{2}{c}{\textbf{Break point UR test}} & -4.6666 &  0.1681 & Oct-07 & -5.4659 & <0.05** & Oct-08 & -4.6934 &  0.1596 & Oct-07 & -5.7877 & ***   & Jun-07 \\
    \multicolumn{14}{c}{\textbf{}} \\
    \hline
    \multicolumn{2}{c}{\textbf{Model Estimation }} & \textbf{est} & \textbf{HAC s.e} & \textbf{p-value} & \textbf{est} & \textbf{HAC s.e} & \textbf{p-value} & \textbf{est} & \textbf{HAC s.e} & \textbf{p-value} & \textbf{est} & \textbf{HAC s.e} & \textbf{p-value} \\
\hline
    \end{tabular}%
  \label{tab:addlabel}%
\end{table}%
\end{landscape}


\newpage

\begin{landscape}

\begin{table}[h!]
  \centering
  \caption{European Financial Stress Indicator CISS level series and its Components}
  \renewcommand\tabcolsep{3pt}
    \begin{tabular}{cccccccccccccc}
    \hline
    \multicolumn{2}{c}{\textbf{Financial Series}} & \multicolumn{3}{c}{\textbf{CISS Index}} & \multicolumn{3}{c}{\textbf{CISS Money Market}} & \multicolumn{3}{c}{\textbf{CISS Financial Market}} & \multicolumn{3}{c}{\textbf{CISS Bond Market}} \\
    \hline
    \multicolumn{2}{c}{\textbf{Unit root testing}} & \textbf{ADF} & \textbf{p-value} & \textbf{Date} & \textbf{ADF} & \textbf{p-value} & \textbf{Date} & \textbf{ADF} & \textbf{p-value} & \textbf{Date} & \textbf{ADF} & \textbf{p-value} & \textbf{Date} \\
    \hline
    \multicolumn{2}{c}{\textbf{Unit root test}} & -2.0761 & 0.5557 & -     & -3.6279 & **    & -     & -2.4794 & 0.3381 & -     & -2.6862 & 0.2435 & - \\
    \multicolumn{2}{c}{\textbf{Break point UR test}} & -4.9578 & 0.0855* & Jul-07 & -5.7272 & ***   & Jun-07 & -4.6060 &  0.1902 & Jun-07 & -4.4741 &  0.2487 & Jun-07 \\
    \multicolumn{14}{c}{\textbf{}} \\
    \hline
    \multicolumn{2}{c}{\textbf{Model Estimation }} & \textbf{est} & \textbf{HAC s.e} & \textbf{p-value} & \textbf{est} & \textbf{HAC s.e} & \textbf{p-value} & \textbf{est} & \textbf{HAC s.e} & \textbf{p-value} & \textbf{est} & \textbf{HAC s.e} & \textbf{p-value} \\
    \hline
    \multirow{2}[2]{*}{Regression model } & $\mu$ & 0.0083 & 0.0031 & ***   & 0.0051 & 0.0012 & ***   & 0.0065 & 0.0021 & ***   & 0.0034 & 0.0015 & *** \\
          & $\rho$ & 0.9577 & 0.0177 & ***   & 0.8816 & 0.0266 & ***   & 0.9435 & 0.0156 & ***   & 0.9325 & 0.0277 & *** \\
\hline
    \end{tabular}%
  \label{tab:addlabel}%
\end{table}%
\end{landscape}


\newpage

\begin{landscape}
\begin{table}[h!]
  \centering
  \caption{CISS Sovereign Stress Indicator - Country level series Indices}
  \renewcommand\tabcolsep{3pt}
    \begin{tabular}{ccccccccccccrc}
    \hline
    \multicolumn{2}{c}{\textbf{Financial Series}} & \multicolumn{3}{c}{\textbf{CISS Sovereign Stress EU}} & \multicolumn{3}{c}{\textbf{CISS Sovereign Germany}} & \multicolumn{3}{c}{\textbf{CISS Sovereign France}} & \multicolumn{3}{c}{\textbf{CISS Sovereign Finland}} \\
    \hline
    \multicolumn{2}{c}{\textbf{Unit root testing}} & \textbf{ADF} & \textbf{p-value} & \textbf{Date} & \textbf{ADF} & \textbf{p-value} & \textbf{Date} & \textbf{ADF} & \textbf{p-value} & \textbf{Date} & \textbf{ADF} & \multicolumn{1}{c}{\textbf{p-value}} & \textbf{Date} \\
    \hline
    \multicolumn{2}{c}{\textbf{Unit root test}} & -0.8486 & 0.9583 & -     & -2.4107 & 0.3728 & -     & -2.4271 & 0.3644 & -     & -2.0766 & \multicolumn{1}{c}{0.5551} & - \\
    \multicolumn{2}{c}{\textbf{Break point UR test}} & -3.4594 & 0.8323 & May-07 & -4.8951 &  0.0997* & May-07 & -4.5519 &  0.2145 & Oct-07 & -4.2863 & \multicolumn{1}{l}{ 0.3444} & Jun-07 \\
    \multicolumn{14}{c}{\textbf{}} \\
    \hline
    \multicolumn{2}{c}{\textbf{Model Estimation }} & \textbf{est} & \textbf{HAC s.e} & \textbf{p-value} & \textbf{est} & \textbf{HAC s.e} & \textbf{p-value} & \textbf{est} & \textbf{HAC s.e} & \textbf{p-value} & \textbf{est} & \multicolumn{1}{c}{\textbf{HAC s.e}} & \textbf{p-value} \\
    \hline
    \multirow{2}[2]{*}{Regression model } & $\mu$ & 0.0034 & 0.0022 & 0.1228 & 0.0054 & 0.0018 & ***   & 0.0056 & 0.0018 & ***   & 0.0058 & \multicolumn{1}{c}{0.0023} & *** \\
          & $\rho$ & 0.9851 & 0.0153 & ***   & 0.9540 & 0.0270 & ***   & 0.9568 & 0.0237 & ***   & 0.9528 & \multicolumn{1}{c}{0.0337} & *** \\
\hline
    \end{tabular}%
  \label{tab:addlabel}%
\end{table}%
\end{landscape}


\newpage

\begin{landscape}

\begin{table}[h!]
  \centering
  \caption{CISS Sovereign Stress Indicator - Country level series Indices (continued)}
  \renewcommand\tabcolsep{3pt}
    \begin{tabular}{rrrrrrrrrrrrrrr}
    \hline
    \multicolumn{3}{c}{\textbf{CISS Sovereign Austria}} & \multicolumn{3}{c}{\textbf{CISS Sovereign Belgium}} & \multicolumn{3}{c}{\textbf{CISS Sovereign Ireland}} & \multicolumn{3}{c}{\textbf{CISS Sovereign Netherlands}} & \multicolumn{3}{c}{\textbf{CISS Sovereign Italy}} \\
    \hline
    \multicolumn{1}{c}{\textbf{ADF}} & \multicolumn{1}{c}{\textbf{p-value}} & \multicolumn{1}{c}{\textbf{Date}} & \multicolumn{1}{c}{\textbf{ADF}} & \multicolumn{1}{c}{\textbf{p-value}} & \multicolumn{1}{c}{\textbf{Date}} & \multicolumn{1}{c}{\textbf{ADF}} & \multicolumn{1}{c}{\textbf{p-value}} & \multicolumn{1}{c}{\textbf{Date}} & \multicolumn{1}{c}{\textbf{ADF}} & \multicolumn{1}{c}{\textbf{p-value}} & \multicolumn{1}{c}{\textbf{Date}} & \multicolumn{1}{c}{\textbf{ADF}} & \multicolumn{1}{c}{\textbf{p-value}} & \multicolumn{1}{c}{\textbf{Date}} \\
    \hline
    \multicolumn{1}{c}{-1.8575} & \multicolumn{1}{c}{ 0.6725} & \multicolumn{1}{c}{-} & \multicolumn{1}{c}{-2.0299} & \multicolumn{1}{c}{0.5809} & \multicolumn{1}{c}{-} & \multicolumn{1}{c}{-1.3917} & \multicolumn{1}{c}{0.8605} & \multicolumn{1}{c}{-} & \multicolumn{1}{c}{-1.9953} & \multicolumn{1}{c}{0.5997} & \multicolumn{1}{c}{-} & \multicolumn{1}{c}{-1.3276} & \multicolumn{1}{c}{0.8779} & \multicolumn{1}{c}{-} \\
    \multicolumn{1}{c}{-4.9239} & \multicolumn{1}{c}{0.0924*} & \multicolumn{1}{c}{Jun-06} & \multicolumn{1}{c}{-4.6714} & \multicolumn{1}{c}{0.1666} & \multicolumn{1}{c}{Jul-12} & \multicolumn{1}{c}{-5.3830} & \multicolumn{1}{c}{**} & \multicolumn{1}{c}{Mar-10} & \multicolumn{1}{c}{-6.5018} & \multicolumn{1}{c}{***} & \multicolumn{1}{c}{Dec-07} & \multicolumn{1}{c}{-4.7801} & \multicolumn{1}{c}{0.1299} & \multicolumn{1}{c}{May-11} \\
          &       &       &       &       &       &       &       &       &       &       &       &       &       &  \\
\hline
    \end{tabular}%
  \label{tab:addlabel}%
\end{table}%
\end{landscape}


\newpage

\begin{landscape}

\begin{table}[h!]
  \centering
  \caption{US Financial Conditions Indicator NFCI first difference series and its components}
  \renewcommand\tabcolsep{3pt}
    \begin{tabular}{cccccccccccccc}
    \hline
    \multicolumn{2}{c}{\textbf{Financial Series}} & \multicolumn{3}{c}{\textbf{NFCI}} & \multicolumn{3}{c}{\textbf{NFCI Risk}} & \multicolumn{3}{c}{\textbf{NFCI Credit}} & \multicolumn{3}{c}{\textbf{NFCI Leverage}} \\
    \hline
    \multicolumn{2}{c}{\textbf{Unit root testing}} & \textbf{ADF} & \textbf{p-value} & \textbf{Date} & \textbf{ADF} & \textbf{p-value} & \textbf{Date} & \textbf{ADF} & \textbf{p-value} & \textbf{Date} & \textbf{ADF} & \textbf{p-value} & \textbf{Date} \\
    \hline
    \multicolumn{2}{c}{\textbf{Unit root test}} & -6.9821 & ***   & -     & -7.6008 & ***   & -     & -6.1752 & ***   & -     & -5.5624 & ***   & - \\
    \multicolumn{2}{c}{\textbf{Break point UR test}} & -10.491 & ***   & Oct-08 & -14.7859 & ***   & Oct-08 & -8.3575 & ***   & Feb-09 & -9.1240 & ***   & Sep-08 \\
    \multicolumn{14}{c}{} \\
    \hline
    \multicolumn{2}{c}{\textbf{Model Estimation }} & \textbf{est} & \textbf{HAC s.e} & \textbf{p-value} & \textbf{est} & \textbf{HAC s.e} & \textbf{p-value} & \textbf{est} & \textbf{HAC s.e} & \textbf{p-value} & \textbf{est} & \textbf{HAC s.e} & \textbf{p-value} \\
    \hline
    \multirow{2}[2]{*}{Regression model } & $\mu$ & -0.0010 & 0.0060 & 0.8632 & -0.0014 & 0.0065 & 0.8262 & 0.0003 & 0.0048 & 0.9463 & -0.0029 & 0.0054 & 0.5874 \\
          & $\rho$ & 0.4044 & 0.0505 & ***   & 0.4279 & 0.0467 & ***   & 0.5102 & 0.0687 & ***   & 0.7043 & 0.0332 & *** \\
\hline
    \end{tabular}%
  \label{tab:addlabel}%
\end{table}%
\end{landscape}

\newpage

\begin{landscape}

\begin{table}[h!]
  \centering
  \caption{US Financial Conditions Indicator NFCI first difference series and its components (sequential monitoring)}
  \renewcommand\tabcolsep{3pt}
    \begin{tabular}{cccccccccccccc}
    \hline
    \multicolumn{2}{c}{\textbf{Financial Series}} & \multicolumn{3}{c}{\textbf{NFCI}} & \multicolumn{3}{c}{\textbf{NFCI Risk}} & \multicolumn{3}{c}{\textbf{NFCI Credit}} & \multicolumn{3}{c}{\textbf{NFCI Leverage}} \\
    \hline
    \multicolumn{2}{c}{\textbf{Sequential Monitoring}} & \multicolumn{12}{c}{} \\
    \hline
    \multicolumn{14}{c}{\textbf{Regression model }} \\
    \hline
    \multicolumn{2}{c}{\textbf{Bai-Perron}} & \multicolumn{2}{c}{\textbf{Criterion}} & \textbf{Date} & \multicolumn{2}{c}{\textbf{Criterion}} & \textbf{Date} & \multicolumn{2}{c}{\textbf{Criterion}} & \textbf{Date} & \multicolumn{2}{c}{\textbf{Criterion}} & \textbf{Date} \\
    \hline
    \multicolumn{2}{c}{Schwarz} & \multicolumn{2}{c}{-3.8495} & Oct-08 & \multicolumn{2}{c}{-3.8197} & Nov-08 & \multicolumn{2}{c}{-4.2349} & Oct-08 & \multicolumn{2}{c}{-3.8917} & Nov-08 \\
    \multicolumn{2}{c}{LWZ } & \multicolumn{2}{c}{-3.7942} &       & \multicolumn{2}{c}{-3.7600} &       & \multicolumn{2}{c}{-4.1796} &       & \multicolumn{2}{c}{-3.8703} &  \\
    \hline
    \multicolumn{2}{c}{\textbf{Standard Process}} & \multicolumn{12}{c}{} \\
    \hline
    \multicolumn{2}{c}{OLS-CUSUM-SQ} & \multicolumn{3}{c}{b3 = Oct 2008} & \multicolumn{3}{c}{b3 = Nov 2008} & \multicolumn{3}{c}{b3 = Oct 2008} & \multicolumn{3}{c}{b3 = Nov 2008} \\
\hline
    \end{tabular}%
  \label{tab:addlabel}%
\end{table}%
\end{landscape}


\begin{landscape}
\begin{table}[h!]
  \centering
  \caption{US Financial Stability Indicators first difference series (St.Louis, Clevland, Kansas City)}
   \renewcommand\tabcolsep{3pt}
    \begin{tabular}{ccccccccccc}
    \hline
    \multicolumn{2}{c}{\textbf{Financial Series}} & \multicolumn{3}{c}{\textbf{STLFSI}} & \multicolumn{3}{c}{\textbf{CFSI}} & \multicolumn{3}{c}{\textbf{KCFSI}} \\
    \hline
    \multicolumn{2}{c}{\textbf{Unit root testing}} & \textbf{ADF} & \textbf{p-value} & \textbf{Date} & \textbf{ADF} & \textbf{p-value} & \textbf{Date} & \textbf{ADF} & \textbf{p-value} & \textbf{Date} \\
    \hline
    \multicolumn{2}{c}{\textbf{Unit root test}} & -11.950 & ***   & -     & -13.289 & ***   & -     & -6.379 & ***   & - \\
    \multicolumn{2}{c}{\textbf{Break point UR test}} & -20.891 & ***   & Oct-08 & -13.558 & ***   & Sep-11 & -19.839 & ***   & Oct-08 \\
    \multicolumn{11}{c}{} \\
    \hline
    \multicolumn{2}{c}{\textbf{Model Estimation }} & \textbf{est} & \textbf{HAC s.e} & \textbf{p-value} & \textbf{est} & \textbf{HAC s.e} & \textbf{p-value} & \textbf{est} & \textbf{HAC s.e} & \textbf{p-value} \\
    \hline
    \multirow{2}[2]{*}{Regression model } & $\mu$ & -0.0049 & 0.0148 & 0.7399 & 0.0040 & 0.0222 & 0.8583 & -0.0042 & 0.0154 & 0.7880 \\
          & $\rho$ & 0.1715 & 0.0351 & ***   & 0.0708 & 0.0765 & 0.3557 & 0.1255 & 0.0426 & *** \\
\hline
    \end{tabular}%
  \label{tab:addlabel}%
\end{table}%

\end{landscape}


\begin{landscape}
\begin{table}[h!]
  \centering
  \caption{US Financial Stability Indicators first difference series (St.Louis, Clevland, Kansas City) (sequential monitoring)}
   \renewcommand\tabcolsep{3pt}
    \begin{tabular}{ccccccccccc}
    \hline
    \multicolumn{2}{c}{\textbf{Financial Series}} & \multicolumn{3}{c}{\textbf{STLFSI}} & \multicolumn{3}{c}{\textbf{CFSI}} & \multicolumn{3}{c}{\textbf{KCFSI}} \\
    \hline
    \multicolumn{2}{c}{\textbf{Sequential Monitoring}} & \multicolumn{9}{c}{} \\
    \hline
    \multicolumn{11}{c}{\textbf{Regression model }} \\
    \hline
    \multicolumn{2}{c}{\textbf{Bai-Perron}} & \multicolumn{2}{c}{\textbf{Criterion}} & \textbf{Date} & \multicolumn{2}{c}{\textbf{Criterion}} & \textbf{Date} & \multicolumn{2}{c}{\textbf{Criterion}} & \textbf{Date} \\
    \multicolumn{2}{c}{Schwarz} & \multicolumn{2}{c}{-2.3375} & Nov-08 & \multicolumn{2}{c}{-1.6254} &       & \multicolumn{2}{c}{-2.0050} & Nov-08 \\
    \multicolumn{2}{c}{LWZ } & \multicolumn{2}{c}{-2.2713} &       & \multicolumn{2}{c}{-1.5700} &       & \multicolumn{2}{c}{-1.9494} &  \\
    \hline
    \multicolumn{2}{c}{\textbf{Standard Process}} & \multicolumn{9}{c}{} \\
    \hline
    \multicolumn{2}{c}{OLS-CUSUM-SQ} & \multicolumn{3}{c}{b3 = Nov 2008} & \multicolumn{3}{c}{b3 = NA} & \multicolumn{3}{c}{b3 = Nov 2008} \\
\hline
    \end{tabular}%
  \label{tab:addlabel}%
\end{table}%

\end{landscape}


\begin{landscape}

\begin{table}[h!]
  \centering
  \caption{US Financial Spread, yield and Volatility Indicators first difference series}
  \renewcommand\tabcolsep{3pt}
    \begin{tabular}{cccccccccccccc}
    \hline
    \multicolumn{2}{c}{\multirow{4}[2]{*}{\textbf{Financial Series}}} & \multicolumn{3}{c}{\textbf{BofA Merrill Lynch US }} &       &       &       & \multicolumn{3}{c}{\textbf{Moody's Seasoned Baa }} &       &       &  \\
    \multicolumn{2}{c}{} & \multicolumn{3}{c}{\textbf{Corporate 3-5 Year }} & \multicolumn{3}{c}{\textbf{Ted Spread}} & \multicolumn{3}{c}{\textbf{Corporate Bond Yield }} & \multicolumn{3}{c}{\textbf{CBOE Volatility Index}} \\
    \multicolumn{2}{c}{} & \multicolumn{3}{c}{\textbf{Option-Adjusted Spread}} & \multicolumn{3}{c}{\textbf{}} & \multicolumn{3}{c}{\textbf{Relative to Yield on 10  Year}} & \multicolumn{3}{c}{\textbf{VIX}} \\
    \multicolumn{2}{c}{} & \multicolumn{3}{c}{}  & \multicolumn{3}{c}{}  & \multicolumn{3}{c}{\textbf{ Treasury Constant Maturity}} & \multicolumn{3}{c}{} \\
    \hline
    \multicolumn{2}{c}{\textbf{Unit root testing}} & \textbf{ADF} & \textbf{p-value} & \textbf{Date} & \textbf{ADF} & \textbf{p-value} & \textbf{Date} & \textbf{ADF} & \textbf{p-value} & \textbf{Date} & \textbf{ADF} & \textbf{p-value} & \textbf{Date} \\
    \hline
    \multicolumn{2}{c}{\textbf{Unit root test}} & -8.495 & ***   & -     & -13.250 & ***   & -     & -8.751 & ***   & -     & -12.012 & ***   & - \\
    \multicolumn{2}{c}{\textbf{Break point UR test}} & -12.213 & ***   & Oct-08 & -14.480 & ***   & Sep-08 & -11.100 & ***   & Oct-08 & -14.744 & ***   & Oct-08 \\
    \multicolumn{14}{c}{} \\
\hline
    \end{tabular}%
  \label{tab:addlabel}%
\end{table}%

\end{landscape}


\begin{landscape}
\begin{table}[h!]
  \centering
  \caption{US Financial Spread, yield and Volatility Indicators first difference series (sequential monitoring)}
   \renewcommand\tabcolsep{3pt}
    \begin{tabular}{cccccccccccccc}
    \hline
    \multicolumn{2}{c}{\multirow{4}[2]{*}{\textbf{Financial Series}}} & \multicolumn{3}{c}{\textbf{BofA Merrill Lynch US }} &       &       &       & \multicolumn{3}{c}{\textbf{Moody's Seasoned Baa }} &       &       &  \\
    \multicolumn{2}{c}{} & \multicolumn{3}{c}{\textbf{Corporate 3-5 Year }} & \multicolumn{3}{c}{\textbf{Ted Spread}} & \multicolumn{3}{c}{\textbf{Corporate Bond Yield }} & \multicolumn{3}{c}{\textbf{CBOE Volatility Index}} \\
    \multicolumn{2}{c}{} & \multicolumn{3}{c}{\textbf{Option-Adjusted Spread}} & \multicolumn{3}{c}{\textbf{}} & \multicolumn{3}{c}{\textbf{Relative to Yield on 10  Year}} & \multicolumn{3}{c}{\textbf{VIX}} \\
    \multicolumn{2}{c}{} & \multicolumn{3}{c}{}  & \multicolumn{3}{c}{}  & \multicolumn{3}{c}{\textbf{ Treasury Constant Maturity}} & \multicolumn{3}{c}{} \\
    \hline
    \multicolumn{2}{c}{\textbf{Sequential Monitoring}} & \multicolumn{12}{c}{} \\
    \hline
    \multicolumn{14}{c}{\textbf{Regression model }} \\
    \hline
    \multicolumn{2}{c}{\textbf{Bai-Perron}} & \multicolumn{2}{c}{\textbf{Criterion}} & \textbf{Date} & \multicolumn{2}{c}{\textbf{Criterion}} & \textbf{Date} & \multicolumn{2}{c}{\textbf{Criterion}} & \textbf{Date} & \multicolumn{2}{c}{\textbf{Criterion}} & \textbf{Date} \\
    \hline
    \multicolumn{2}{c}{Schwarz} & \multicolumn{2}{c}{-2.9914} & Nov-08 & \multicolumn{2}{c}{-3.0800} & Nov-08 & \multicolumn{2}{c}{-3.3475} & Nov-08 & \multicolumn{2}{c}{2.8250} & Nov-08 \\
\hline
    \end{tabular}%
  \label{tab:addlabel}%
\end{table}%

\end{landscape}


\begin{landscape}
\begin{table}[h!]
  \centering
  \caption{European Financial Stress Indicator CISS first difference series and its Components}
  \renewcommand\tabcolsep{3pt}
    \begin{tabular}{cccccccccccccc}
    \hline
    \multicolumn{2}{c}{\textbf{Financial Series}} & \multicolumn{3}{c}{\textbf{CISS Index}} & \multicolumn{3}{c}{\textbf{CISS Money Market}} & \multicolumn{3}{c}{\textbf{CISS Financial Market}} & \multicolumn{3}{c}{\textbf{CISS Bond Market}} \\
    \hline
    \multicolumn{2}{c}{\textbf{Unit root testing}} & \textbf{ADF} & \textbf{p-value} & \textbf{Date} & \textbf{ADF} & \textbf{p-value} & \textbf{Date} & \textbf{ADF} & \textbf{p-value} & \textbf{Date} & \textbf{ADF} & \textbf{p-value} & \textbf{Date} \\
    \hline
    \multicolumn{2}{c}{\textbf{Unit root test}} & -12.729 & ***   & -     & -12.899 & ***   & -     & -16.520 & ***   & -     & -15.05381 & ***   & - \\
    \multicolumn{2}{c}{\textbf{Break point UR test}} & -14.238 & ***   & Oct-08 & -14.511 & ***   & Oct-08 & -17.152 & ***   & Aug-07 & -15.729 & ***   & May-10 \\
    \multicolumn{14}{c}{} \\
    \hline
    \multicolumn{2}{c}{\textbf{Model Estimation }} & \textbf{est} & \textbf{HAC s.e} & \textbf{p-value} & \textbf{est} & \textbf{HAC s.e} & \textbf{p-value} & \textbf{est} & \textbf{HAC s.e} & \textbf{p-value} & \textbf{est} & \textbf{HAC s.e} & \textbf{p-value} \\
    \hline
    \multirow{2}[2]{*}{Regression model } & $\mu$ & -0.0003 & 0.0027 & 0.9138 & -0.0001 & 0.0004 & 0.7668 & 0.0000 & 0.0013 & 0.9866 & 0.0000 & 0.0006 & 0.9834 \\
          & $\rho$ & 0.1139 & 0.0731 & 0.1206 & -0.1154 & 0.0722 & 0.1114 & -0.1472 & 0.0502 & <0.01* & -0.0548 & 0.0779 & 0.4823 \\
\hline
    \end{tabular}%
  \label{tab:addlabel}%
\end{table}%

\end{landscape}


\begin{landscape}

\begin{table}[h!]
  \centering
  \caption{CISS Sovereign Stress Indicator - Country first difference series Indices}
  \renewcommand\tabcolsep{3pt}
    \begin{tabular}{cccccccccccccc}
    \hline
    \multicolumn{2}{c}{\textbf{Financial Series}} & \multicolumn{3}{c}{\textbf{CISS Sovereign Stress EU}} & \multicolumn{3}{c}{\textbf{CISS Sovereign Germany}} & \multicolumn{3}{c}{\textbf{CISS Sovereign France}} & \multicolumn{3}{c}{\textbf{CISS Sovereign Finland}} \\
    \hline
    \multicolumn{2}{c}{\textbf{Unit root testing}} & \textbf{ADF} & \textbf{p-value} & \textbf{Date} & \textbf{ADF} & \textbf{p-value} & \textbf{Date} & \textbf{ADF} & \textbf{p-value} & \textbf{Date} & \textbf{ADF} & \textbf{p-value} & \textbf{Date} \\
    \hline
    \multicolumn{2}{c}{\textbf{Unit root test}} & -12.414 & ***   & -     & -11.863 & ***   & -     & -11.407 & ***   &       & -14.062 & ***   & - \\
    \multicolumn{2}{c}{\textbf{Break point UR test}} & -13.502 & ***   & Nov-11 & -12.787 & ***   & Mar-08 & -13.467 & ***   & Oct-08 & 16.294 & ***   & Sep-08 \\
    \multicolumn{14}{c}{} \\
    \hline
    \multicolumn{2}{c}{\textbf{Model Estimation }} & \textbf{est} & \textbf{HAC s.e} & \textbf{p-value} & \textbf{est} & \textbf{HAC s.e} & \textbf{p-value} & \textbf{est} & \textbf{HAC s.e} & \textbf{p-value} & \textbf{est} & \textbf{HAC s.e} & \textbf{p-value} \\
    \hline
    \multirow{2}[2]{*}{Regression model } & $\mu$ & 0.000261 & 0.00257 & 0.9193 & -0.0003 & 0.0022 & 0.8804 & -0.0003 & 0.0022 & 0.8865 & 0.0000 & 0.0019 & 0.9845 \\
          & $\rho$ & 0.095331 & 0.07218 & 0.1882 & 0.1364 & 0.0766 & 0.0764 & 0.1746 & 0.0938 & 0.0642 & -0.0312 & 0.0883 & 0.7247 \\
\bottomrule
    \end{tabular}%
  \label{tab:addlabel}%
\end{table}%

\end{landscape}

\newpage   

\begin{wrap}

\subsection{Review of Structural Break Testing Methodologies}

\subsection{Retrospective Structural Break Testing}

When modeling financial time series, one of the initial concerns is the validity of the stationarity assumption (parameter constancy over time) due to the uniqueness property of a stochastic process; while a violation of this assumption might be due to either changes in the mean or the variance at some point in the sample. Therefore, when dealing with financial data appropriate transformations can be used to impose the stationarity assumption (e.g., lag differences) while a useful tool to assess the nature of non-stationarity (i.e., deterministic or stochastic) is to test for unit roots. In particular, \cite{phillips1988testing} proposed a procedure for testing for the presence of a unit root in general time series (e.g., model with drift and linear trend) with regression statistics which have asymptotic properties approximated with the Brownian motion.    


For simulations and estimations purposes we consider a linear regression model with an AR(1) independent variable. The model is given by
\begin{align}
y_t = \mu + \rho y_{t-1} + \epsilon_t, \ t=\{1,...,n,n+1,...,N \} \ \text{and} \ \ \epsilon_t \sim N(0,1) 
\label{main model}
\end{align}
If $\rho=1$ then a unit root occurs which is not desirable since it affects the properties of the model and a unit root testing, tests the null hypothesis $H_0: \rho_t =1 \forall t$. Moreover, model specification is another issue of concern and in particular  \cite{costantini2016simple} developed a testing procedure for both unit root and model misspecification applied on real exchange interest rates; time series with a single structural break.

\subsection{historical structural break testing }

Using the sequential break point detection processes for monitoring financial time series, first requires to monitor the full sample (and historical period) in order to obtain information regarding the existence of breaks, the number of breaks and the location of breaks. Different approach is required depending on the number of breaks. 

In particular, \citep{bai1994least} examines the case of a single mean shift in a linear time series process and estimates the unknown shift point by the method of least squares. The consistency and rate of convergence of the estimated change point along with the asymptotic distribution of the change point estimator are examined. The variance of the change point estimator (seen by the ARL in simulations) is shown to be affected by serial correlation, inherited through the sum of the coefficients of the linear process.    

\begin{definition}(Non-sequential Break point estimator) Consider the case of a single break in the time series given by a simplified version of the econometrics model \ref{eq:model}   
\begin{align*}
 y_t &= \mu_1 + X_t, \ \text{if} \ \ t \leq \kappa^{0} \\
 y_t &= \mu_2 + X_t, \ \text{if} \ \ \kappa^{0} + 1 \leq t \leq N
\end{align*}
where $\mu_i$ is the mean of regime $i,  i=1,2$ and $X_t$ is a linear process of martingale differences and $\kappa^{0}$ is the unknown break point. Let the full sample have finite many different splits and denote $\bar{y}_k$ the mean of the first $k$ observations and $\bar{y}^{*}_k$ the mean of the last $T-k$ observations. We define the sum of squared residuals to be
\begin{align}
 S_N(k) = \sum_{t=1}^k (y_t - \bar{y}_k)^2 + \sum_{t=k+1}^N (y_t - \bar{y}^{*}_k)^2
\end{align}    
then the non-sequential break point estimator is defined as $\hat{\kappa} = \text{argmin}_{[1 \leq t \leq N-1]} \ S_N(k).$
\end{definition}

The examination for multiple structural breaks is proposed by \citep{bai1997estimating} and is done using the method of sequential computation of break points one at a time by repartitioning the full sample in order to get consistent estimators. Moreover, \cite{bai1998estimating} study the properties of the break point estimator and its limiting distribution for the multiple structural breaks case. \cite{bai2003computation} present a dynamic programming algorithm with which partitions are sequentially examined whether they achieve a global minimization of the overall sum of squared residuals. 

Let $N$ be the sample size, $\nu$ the number of breaks, $h$ the minimum distance between breaks and $SSR(i,j)$ the sum of squared residuals obtained from a segment in the time interval $[i,j].$ Then $SSR(\{\psi_{\nu,f}\})$ is the sum of squared residuals associated with the optimal partition of $\nu$ breaks and $f$ first observations. The optimal partition solves the following recursive problem \citep{bai2003computation}:
\begin{align}
SSR(\{\psi_{\nu,f}\}) = \mathsf{min}_{\nu h \leq j \leq N-h} \big[ SSR(\{\psi_{\nu-1,j}\}) + SSR(\{\psi_{j+1,N}\}) \big]
\end{align}

\end{wrap}

\newpage 

\section{Asymptotic Theory}
\label{AppendixD}

The OLS-CUSUM test (\cite{kramer1988testing}) belongs to the class of residual based statistics (see, \cite{stock1994unit}) based on the partial sum process of regression residuals. We define a general class of regression residuals based on the partial sum process (see, \cite{kulperger2005high}). 

\begin{definition}\label{definition1}
The $m-$th order moment partial sum process of residuals is given by 
\begin{align*}
\widehat{S}_n^{(m)}(r) =  \sum_{t=1}^{[nr]} \widehat{\epsilon}^{(m)}_t , \ 0 \leq r \leq 1, m \in \mathbb{Z} \ \text{with} \ m \geq 1.
\end{align*}
with the partial sum process of the corresponding innovations is similarly defined as
\begin{align*}
S_n^{(m)}(r) =  \sum_{t=1}^{[nr]} \epsilon^{(m)}_t , \ 0 \leq r \leq 1, \ 0 \leq r \leq 1, m \in \mathbb{Z} \ \text{with} \ m \geq 1.
\end{align*}
\end{definition}

\begin{theorem}\label{Theorem1}
Let $m \geq 1$ be an integer and $\sqrt{n}| \hat{ \theta}_n - \theta  | = \mathcal{O}_p(1)$ where $\hat{ \theta}_n$ is the set of estimated model parameters and $\theta$ is in the interior of $\Theta$. If $\mathbb{E} \left( | \epsilon_0|^m \right) < \infty$, then 
\begin{align}
\underset{ r \in [0,1] }{  \text{sup} } \frac{1}{ \sqrt{n} } \left| \left(  \widehat{S}_n^{(m)}(r) - r \widehat{S}_n^{(m)}(r) \right) - \left(  S_n^{(m)}(r) - r S_n^{(m)}(r) \right) \right|  = o_p(1).
\end{align}
Then, the invariance principle for partial sums for an i.i.d sequence $\{ \epsilon_t^{(m)} \}$ implies that 
\begin{align}
\left\{  \frac{ S_n^{(m)}(r) - r S_n^{(m)}(r) }{ \sigma_m \sqrt{n} }, 0 \leq r \leq 1 \right\}  \ \text{and} \  \left\{  \frac{ \widehat{S}_n^{(m)}(r) - r \widehat{S}_n^{(m)}(r) }{ \sigma_m \sqrt{n} }, 0 \leq r \leq 1 \right\} 
\end{align} 
both converge weakly in the Skorokhod space $D[0,1]$ to a Brownian bridge $\{ G(r), 0 \leq r \leq 1 \}$.
\end{theorem}
We focus on the cases where $m=1$ and $m=2$ which represent functionals of the OLS-CUSUM and OLS-CUSUM squared (see e.g., \cite{deng2008limit}). Furthermore, we omit the proof of Theorem \ref{Theorem1}, which demonstrates a weak invariance principle, a stronger version of Donsker's classical functional central limit theorem (see, e.g., \cite{kulperger2005high} and \citep{csorgHo2003donsker}). The limit theory related to the residual based tests and Wald type statistics for detecting structural change is based on the implications of Theorem \ref{Theorem1}. In this paper, we focus on functionals of the OLS-CUSUM test and their related limit theory. The weakly convergence of the asymptotic distribution of the OLS-CUSUM statistic in the case of the classical regression model is examined by \cite{aue2013structural}. 
\begin{example}\label{Example2}
Consider the following model 
\begin{align}
y_t = x_t^{\prime} \beta_1 \mathbf{1} \{ t \leq k \}  + x_t^{\prime} \beta_2 \mathbf{1} \{ t > k \} + \epsilon_t, \ t =1,...,n  
\end{align}
where $y_t$ is the regressand, $x_t = [1, x_{2,t},...,x_{K,t} ]^{\prime} = [1, \tilde{x}_t^{\prime} ]^{\prime}$ is a $K-$dimensional vector of regressors (including an intercept) and $\epsilon_t$ are i.i.d $(0, \sigma_\epsilon^2)$ innovations. Define $x_{1,t} \equiv   x_t^{\prime} \mathbf{1} \{ t \leq k \}$ and $x_{2,t} \equiv   x_t^{\prime} \mathbf{1} \{ t > k \}$ for $k = [nr]$ with $r \in [0,1]$. Under the null hypothesis of no structural break $\mathbb{H}_0: \beta_1 = \beta_2 \equiv \beta$. Moreover, $\hat{\beta}_n$ is the $\sqrt{n}$-consistent estimator of $\beta$ such that $\sqrt{n} \left( \hat{\beta}_n - \beta \right) = \mathcal{O}_p(1)$. Then, the OLS-CUSUM statistic, $C_n(k)$, is constructed based on the OLS residuals under the null hypothesis, given by $\hat{\epsilon_t} = y_t - \hat{\beta}_n x_t = \epsilon_t - x_t^{\prime} \left( \hat{\beta}_n - \beta  \right)$.

\newpage

Therefore, we have that  
\begin{align}
C_n(k) = \frac{1}{\hat{\sigma_{\epsilon}} } \left(   \frac{1}{\sqrt{n}} \sum_{t=1}^{[nr]} \hat{\epsilon_t}      - \frac{[nr]}{n}  \frac{1}{\sqrt{n}} \sum_{t=1}^n \hat{\epsilon_t}  \right)
\end{align}
First, the OLS residuals can be expressed as below
\begin{align}\label{OLSresiduals}
\frac{1}{\sqrt{n} } \sum_{t=1}^{ [nr] }  \hat{ \epsilon}_t = \frac{1}{\sqrt{n} } \sum_{t=1}^{ [nr] } \epsilon_t - \frac{1}{\sqrt{n} }  \sum_{t=1}^{ [nr] }  x_t^{\prime} \left( \hat{\beta}_n - \beta  \right)
\end{align}
Second, the following asymptotic result holds  
\begin{align}\label{PM1992}
\frac{1}{\sqrt{n}} \sum_{t=1}^{ [nr] } x_t^{\prime} \left( \hat{\beta}_n - \beta  \right) = \frac{r}{\sqrt{n}} \sum_{t=1}^n \epsilon_t + o_p(1).
\end{align}
A short proof on the asymptotic result above is provided here. We can express the left side of (\ref{PM1992}) as an inner product since our framework allows such representation
\begin{align}
\left[ \frac{1}{\sqrt{n}} \sum_{t=1}^{ [nr] } x_t^{\prime} \left( \hat{\beta}_n - \beta  \right) \right] = \left[  \frac{1}{n} \sum_{t=1}^{[nr]} x_t^{\prime} \right] . \left[ \sqrt{n} \left( \hat{\beta}_n - \beta  \right) \right]
\end{align}
The first term $\left(  \frac{1}{n} \sum_{t=1}^{[nr]} x_t^{\prime} \right) \overset{p}{\to} [r, 0,...,0]$ since    $\underset{ n \to \infty }{ \text{lim}} \frac{1}{n} \sum_{t=1}^{[nr]} \tilde{x}_t = 0$ and $\underset{ n \to \infty }{ \text{lim}} \frac{1}{n} \sum_{t=1}^{[nr]} 1 = r$. The second term by considering a matrix decomposition for $Q = \left( \frac{1}{n} \sum_{t=1}^n x_t x_t^{\prime} \right)$ where $x_t=[1, \tilde{x}_t^{\prime} ]^{\prime}$ can be expressed as following. 
\begin{align}
\sqrt{n} \left( \hat{\beta}_n - \beta  \right)  = \left( \frac{1}{n} \sum_{t=1}^n x_t x_t^{\prime} \right)^{-1} \left( \frac{1}{\sqrt{n}} \sum_{t=1}^n x_t \epsilon_t  \right)  =
\frac{1}{ \sqrt{n} }
\begin{bmatrix}
1 & \underline{0} \\
\underline{0} & \tilde{Q}
\end{bmatrix}^{-1} 
\begin{bmatrix}
\sum_{t=1}^n  \epsilon_t \\
\sum_{t=1}^n  \tilde{x}_t \epsilon_t
\end{bmatrix} + o_p(1)
\end{align}
since $\underset{ n \to \infty }{ \text{lim}} \frac{1}{n} \sum_{t=1}^{[nr]} \tilde{x}_t \tilde{x}_t^{ \prime} = \tilde{Q}$. Also, note that $\begin{bmatrix}
1 & \underline{0} \\
\underline{0} & \tilde{Q}
\end{bmatrix}^{-1} = \begin{bmatrix}
1 & \underline{0} \\
\underline{0} & \tilde{Q}^{-1}
\end{bmatrix}$. 
Therefore, we obtain
\begin{small}
\begin{align}
\left[ \frac{1}{\sqrt{n}} \sum_{t=1}^{ [nr] } x_t^{\prime} \left( \hat{\beta}_n - \beta  \right) \right]  \overset{p}{\to} [r \  \underline{0}] \frac{1}{ \sqrt{n} }  \begin{bmatrix}
1 & \underline{0} \\
\underline{0} & \tilde{Q}^{-1}
\end{bmatrix} \begin{bmatrix}
\sum_{t=1}^n  \epsilon_t \\
\sum_{t=1}^n  \tilde{x}_t \epsilon_t
\end{bmatrix} 
= \frac{r}{\sqrt{n}} \sum_{t=1}^n \epsilon_t + o_p(1)
\end{align}
\end{small}
where $\underline{0}$ is $(K-1)$ dimensional column vectors of zeros. Using the limit given by (\ref{PM1992}) and the expression for the OLS residuals (\ref{OLSresiduals}) the OLS-CUSUM statistic has the following formulation
\begin{align*}
C_n(k) &= \frac{1}{\hat{\sigma_{\epsilon}} } \frac{1}{ \sqrt{n} } \left\{  \left(   \sum_{t=1}^{k} \epsilon_t -       \sum_{t=1}^{k} x_t^{\prime} \left( \hat{\beta}_n - \beta  \right) \right) - r \left(   \sum_{t=1}^{n} \epsilon_t -       \sum_{t=1}^{n} x_t^{\prime} \left( \hat{\beta}_n - \beta  \right) \right) \right\} \\
&= \frac{1}{\hat{\sigma_{\epsilon}} } \frac{1}{ \sqrt{n} } \left\{  \left(   \sum_{t=1}^{k} \epsilon_t -        r \sum_{t=1}^{n} \epsilon_t \right) - \left(  \sum_{t=1}^{k} x_t^{\prime} \left( \hat{\beta}_n - \beta  \right) -       r \sum_{t=1}^{n} x_t^{\prime} \left( \hat{\beta}_n - \beta  \right) \right) \right\}
\end{align*}
Since the second term above gives that $r \left( \sum_{t=1}^n \epsilon_t - \sum_{t=1}^n \hat{\epsilon}_t \right) = o_p(1)$ then the result follows, 
\begin{align}
C_n(k) = \underset{ r \in [\nu , 1 - \nu]  }{ \text{sup} }  \left\{ \frac{1}{\hat{\sigma_{\epsilon}} } \left(   \sum_{t=1}^{k} \frac{ \epsilon_t }{ \sqrt{n} } -   r \sum_{t=1}^{n} { \epsilon_t }{  \sqrt{n} } \right) \right\} \Rightarrow [ W(r) - r W(1) ], r \in [0,1].
\end{align} 
showing that $C_n(k)$ weakly converges to the Brownian bridge uniformly for $r \in [0,1]$.
\end{example}

\newpage

\end{document}